\documentclass[aps,pra,floatfix,amsmath,amssymb,superscriptaddress,showpacs,showkeys,onecolumn,10pt]{revtex4-2}
\usepackage[caption=false]{subfig}
\usepackage{graphicx,bm,color}
\usepackage{amsfonts}
\usepackage{color}
\usepackage{algorithm2e}
\usepackage{braket}
\usepackage[dvipsnames]{xcolor}
\usepackage{hyperref}
\hypersetup{
    colorlinks=true,
    linkcolor=blue,
    filecolor=blue,      
    urlcolor=blue,
}
\usepackage{epstopdf}

\begin{document}

\frenchspacing

\title{A new error-modeling of Hardy's paradox for superconducting qubits and its experimental verification}
\author{Soumya Das}
\email{soumya06.das@gmail.com}
\affiliation{Cryptology and Security Research Unit, R. C. Bose Centre for Cryptology and Security, Indian Statistical Institute, Kolkata 700108, India}
\author{Goutam Paul}
\email{goutam.paul@isical.ac.in}
\affiliation{Cryptology and Security Research Unit, R. C. Bose Centre for Cryptology and Security, Indian Statistical Institute, Kolkata 700108, India}

\begin{abstract}
Hardy's paradox (equivalently, Hardy's non-locality or Hardy's test)  [\href{https://link.aps.org/doi/10.1103/PhysRevLett.68.2981}{L. Hardy, Phys. Rev. Lett. \textbf{68}, 2981 (1992)}] is used to show non-locality without inequalities and it has been tested several times using optical circuits. We, for the first time, experimentally test Hardy's paradox of non-locality in superconducting qubits. 
For practical verification of Hardy's paradox, we argue that the error-modeling used in optical circuits is not useful for superconducting qubits. So, we propose a new error-modeling for Hardy's paradox and a new method to estimate the lower bound on Hardy's probability (i.e., the probability of a specific event in Hardy's test) for superconducting qubits. Our results confirmed the theory that any non-maximally entangled state of two qubits violates Hardy's equations; whereas, any maximally entangled state and product state of two qubits do not exhibit Hardy's non-locality. Further, we point out the difficulties associated with the practical implementation of quantum protocols based on Hardy's paradox  and propose possible remedies. We also propose two performance measures for any two qubits of any quantum computer based on superconducting qubits.
\end{abstract}


\maketitle

\section{Introduction} \label{sec:1}
In 1935, the Einstein-Podolsky-Rosen (EPR) paradox~\cite{epr} raised the question about the completeness of the quantum theory and  claimed that  nature should be  described by any local-realistic theory, also known as local hidden variable (LHV) theory~\cite{Genovese05}. In 1964, the non-local characteristic of quantum theory was established by Bell's theorem~\cite{bell}. A revised version of Bell's theorem, known as the Clause-Horne-Shimony-Holt (CHSH) inequality~\cite{CHSH69}, is used mostly for experiments to test non-locality. A violation of the CHSH inequality in an experiment rules out all the possibilities of describing that experiment with any LHV theory. So far, a significant number of experiments have been performed, demonstrating a violation of the CHSH inequality~\cite{aspect1982,Weihs98,Groblacher07,Salart08,Ansmann09,Guistina13,Larsson14,Christensen13,Lynden15,Hensen15,Giustina15,handsteiner17,Rosenfeld2017}. For this reason, Bell-type inequalities are used as standard tools to differentiate between quantum and classical correlations. A comprehensive review of Bell's theorem including its theoretical and experimental aspects can be found in~\cite{Brunner14}.

Rather than using the statistical inequalities, the contradiction between quantum theory and any LHV theory can also be demonstrated by a simple and elegant way, which is known as all-versus-nothing (AVN) proof of Bell's nonlocality~\cite{Mermin_poly90}. In this method, a logical paradox is formed in such a way that while doing experiments, in principle, only a single event can be used to reveal the non-locality. Among several AVN proofs, in~\cite{GHZ}, the authors have demonstrated non-locality without using inequalities which is known as the Greenberger-Horne-Zeilinger (GHZ) paradox. It has been verified experimentally in~\cite{Pan2000}. However, this paradox applies to three~\cite{GHZ} or more qubits~\cite{Brunner14}, but not for two qubits. In 1992, through a thought experiment, Hardy constructed the test of local realism without using inequalities for two qubits, which is called Hardy's test~\cite{hardy92,hardy93}. It is known as the ``Best version of Bell's theorem" as indicated by Mermin~\cite{mermin}. This test provides a direct contradiction between the predictions of quantum theory and any LHV theory for two qubits~\cite{hardy92,hardy93} and also for multi-qubits~\cite{Cereceda04}.

Several experiments have been performed to demonstrate Hardy's paradox using polarization, energy-time and orbital angular momentum of photons, entangled qubits, classical light, and two-level quantum states~\cite{Giuseppe1997,irvine2005,Lundeen2009,Yokota2009,vallone2011,chen2012,karimi2014,zhang2016,fan2017,ladder2017}, but none in superconducting qubits. The applications of Hardy's paradox includes device-independent randomness~\cite{ramij_random}, device-independent quantum key distribution~\cite{DIQKD}, quantum Byzantine agreement (QBA)~\cite{QBA}, etc. 

\subsection{The IBM quantum computer and significant experiments performed therein}
IBM has given access to its quantum computer that uses superconducting qubits in the cloud and this opens a new door for testing quantum phenomena to the researchers~\cite{IBM}. The CHSH inequality and the GHZ paradox are already performed in the IBM quantum computer~\cite{IBM}.

In~\cite{devitt16}, the author has implemented some protocols in quantum error correction, quantum arithmetic, quantum graph theory, and fault-tolerant quantum computation  in the IBM 
 quantum computer. In~\cite{berta2016}, the authors have tested the theoretical predictions of entropic uncertainty relation with quantum side information (EUR-QSI) in the IBM quantum computer. Leggett-Garg test~\cite{huffman2017}, compressed quantum computation~\cite{alsina2017}, 
fault-tolerant state preparation~\cite{Takita17}, fault-tolerant logical gates~\cite{Harper19}, quantum cheque~\cite{Qcheque}, quantum permutation algorithm~\cite{Yalcinkaya17}, Deutsch-Jozsa-like algorithm~\cite{DJALGOPani18}, Shor's factoring algorithm~\cite{Amico19}, hybrid quantum linear equation algorithm~\cite{Lee19} are also recently performed in the IBM quantum computer. 

\subsection{Motivation and Contributions}
Though Mermin inequalities~\cite{Mermin_poly90} have been tested experimentally using photons and ion traps~\cite{Zhao03,Lanyon14}, subsequently the authors of~\cite{mermin2016} have tested Mermin polynomial of three, four and five qubits in the IBM quantum computer based on superconducting qubits. However, none of the experimental verifications of Hardy's non-locality \cite{Giuseppe1997,irvine2005,Lundeen2009,Yokota2009,vallone2011,chen2012,karimi2014,zhang2016,fan2017,ladder2017,Luo2018,Yang19}
have used superconducting qubits. This motivates us to test Hardy's paradox for two qubits in a quantum computer using superconducting qubits. 

Our contributions in this work can be summarized as follows.
\begin{itemize}
\item We argue that for practical verification of Hardy's test, the error-modeling used for optical circuits cannot be used for superconducting qubits. We propose a new error-modeling and a new method to estimate the lower bound on Hardy's probability for superconducting qubits. 
\item We also point out that the earlier tests performed in optical circuits and in the IBM quantum computer have not analyzed the test results in a statistically correct and coherent way. We analyze our data using Student's t-distribution~\cite{feller} which is the statistically correct way to represent the test results. 
\item We experimentally verify Hardy's paradox for two qubits on a quantum computer based on superconducting circuits. Our statistical analysis leads to the conclusion that any two-qubit non-maximally entangled state (NMES) gives a nonzero value of Hardy's probability, whereas any two-qubit maximally entangled state (MES) as well as any product state (PS) yields a zero value of Hardy's probability.
\item We identify the difficulties associated with the practical implementation  of quantum protocols based on Hardy's paradox  and discuss how to overcome them.
\item We propose two performance measures for any two qubits of any quantum computer based on superconducting qubits.
\end{itemize}

\section{Hardy's test of non-locality} \label{Hardy_test}
Hardy's test of non-locality for two qubits involves two non-communicating distant parties, Alice and Bob. A physical system consisting of two subsystems is shared between them. Alice and Bob can freely measure and observe the measurement results of their own subsystems. Alice can perform the measurement on her own subsystem by choosing freely one of the two $\lbrace+1,-1\rbrace$-valued random variables $A_{1}$ and $A_{2}$. Similarly, Bob can also choose freely one of the two $\lbrace+1,-1\rbrace$-valued random variables $B_{1}$ and $B_{2}$ for measuring the subsystem in his possession. 

Hardy's test of non-locality starts with the following set of joint probability equations.
\begin{align}
P(+1,+1|A_{1},B_{1})=&0,\label{eq:1}\\
P(+1,-1|A_{2},B_{1})=&0,\label{eq:2}\\
P(-1,+1|A_{1},B_{2})=&0,\label{eq:3}\\
P(+1,+1|A_{2},B_{2})=&q, \hspace{0.2cm} \text{where} \hspace{0.2cm} 
  \begin{cases}
       q=0 & \text{for LHV theory,} \label{eq:4}\\
       q>0& \text{for non-locality.} 
  \end{cases}
\end{align}

\noindent Here $P(x,y|A_{i},B_{j})$ denotes the joint probability of obtaining outcomes $x , y \in \left\lbrace +1,-1 \right\rbrace $ given that $ A_{i} $ and $ B_{j} $ were the experimental choices made where $i , j \in \left\lbrace 1 , 2 \right\rbrace $. If an experiment is designed in such a way that Equations~(\ref{eq:1}),~(\ref{eq:2}), and~(\ref{eq:3}) are satisfied, then for any LHV theory, the right-hand side of Equation~(\ref{eq:4}) becomes zero. But if this value is found to be greater than zero for some values of $q$, then non-locality is established. The set of Equations~(\ref{eq:1})-(\ref{eq:4}) are called \textit{Hardy's equations} and $q$ is called \textit{Hardy's probability}. 

It can be easily shown that a greater than zero value of Equation~(\ref{eq:4}) implies non-locality under the assumptions of Equations~(\ref{eq:1})-(\ref{eq:3}). Let us assume that, when $A_{2}$ and $B_{2}$ are measured simultaneously, an event with $A_{2}=B_{2}=+1$ is detected. It can be seen from Equation~(\ref{eq:2}) that the measurement of $B_{1}$ must yield the output +1, as $A_{2}=1$ can never occur with $B_{1}=-1$, the only option left is $B_{1}=+1$. Following the same logic, from Equation~(\ref{eq:3}), it can be concluded that $A_{1}=+1$ must occur, as the value of $B_{2}=1$ can never be detected with $A_{1}=-1$. Also, because of the locality assumption, the value of $B_{1}$  must be independent of whether Alice measures $A_{1}$ or $A_{2}$. Similarly, the value of $A_{1}$  must be independent of whether Bob measures $B_{1}$ or $B_{2}$. So, it can be concluded from LHV theory that the values of $A_{1}$ and $B_{1}$ must be $+1$. But this is not possible as shown in Equation~(\ref{eq:1}). So, given Equations~(\ref{eq:1})-(\ref{eq:3}) are satisfied, a single occurrence of the event $A_{2}=B_{2}=+1$ can rule out all possibilities that experiment can be described by an LHV theory.

The maximum value of Hardy's probability $ q $ is found to be $ q_{max}=\frac{5\sqrt{5}-11}{2} \approx 0.09017  $ for two qubits~\cite{hardy93,rabelo12}. For the two-qubit system, no MES as well as no PS obey Hardy's non-locality, but all NMES exhibit Hardy's non-locality~\cite{gold94}. This is the specialty of Hardy's test that only a single event can discard all LHV theories. The motivation of this work is to validate this statement for the quantum computer using superconducting qubits. As every MES of three or higher qubits exhibits Hardy's non-locality, we restrict our discussion for two qubits only. 

\section{Practical verification of Hardy's test} \label{Practical Hardy's test} 
For any experimental set-up, it is quite obvious that the joint probabilities described in Equations~(\ref{eq:1})-(\ref{eq:4}) may not be zero due to errors caused by any external environment or internal device or both. So, Equations~(\ref{eq:1})-(\ref{eq:4}) can be written with some error parameter $\epsilon$~\cite{ghirardi06}.
Here we present the error-model in a slightly different manner so that the result of the practical experiment on an unknown state can be interpreted in a statistically correct and coherent way.
\begin{align}
P(+1,+1|A_{1},B_{1})= &\epsilon_{1},\label{eq:5}\\
P(+1,-1|A_{2},B_{1})=&\epsilon_{2},\label{eq:6}\\
P(-1,+1|A_{1},B_{2})=&\epsilon_{3},\label{eq:7}\\
P(+1,+1|A_{2},B_{2})=&\epsilon_{5}=\epsilon_{4}+ q,  \hspace{0.2cm}\text{where} \hspace{0.2cm}
\begin{cases}
       q = 0 & \text{for LHV theory,} \label{eq:8}\\
       q > 0 & \text{for non-locality,}
  \end{cases}
\end{align}
 and $0 \leq \epsilon_{i} \leq 1$, $ \forall$ $ i \in \left\lbrace 1,2,3,5\right\rbrace$.  The bounds of $\epsilon_{4}$ become  $0 \leq \left( \epsilon_{4}+q \right) \leq 1$ or $-q \leq \epsilon_{4} \leq \left( 1-q\right) $. For every MES and every PS of two qubits, the right-hand side of Equation~(\ref{eq:8}) is $\epsilon_{5}=\epsilon_{4}$, i.e., $q=0$, which supports LHV theory. But for every NMES, it is $\epsilon_{5}=\epsilon_{4}+q$ where $q>0$, which supports non-locality.
Thus, by inspecting the values of $q$ in an experiment, it may be possible to infer whether the underlying state is MES/PS or NMES.

\subsection{Connection to the CHSH Inequality}
Using simple set-theoretic arguments, one can show that Hardy's equations are a special case of the famous CHSH inequality~\cite{CHSH69}. The CHSH version of Hardy's Equations~\cite{barun08} is described as
\begin{equation}
 \begin{aligned}
 \label{CHSH}
 P(+1,+1|A_{2},B_{2}) -P(+1,+1|A_{1},B_{1})
- P(+1,-1|A_{2},B_{1})  - P(-1,+1|A_{1},B_{2}) \leq 0.
 \end{aligned}
 \end{equation} 
A violation of Equation~(\ref{CHSH}) means a violation of local realism, which supports non-locality. 
Putting the ideal values of the probabilities from Equations~(\ref{eq:1})-(\ref{eq:4}) into Equation~(\ref{CHSH}),  
we get $q \leq 0$. So, $q=0$ supports LHV theory and $q>0$ supports non-locality.
 But when the practical values of the probabilities from Equations~(\ref{eq:5})-(\ref{eq:8}) are put into Equation~(\ref{CHSH}), we get
\begin{equation} \label{CHSH_noise_2}
\epsilon_{5} - \epsilon_{1}- \epsilon_{2} -\epsilon_{3} \leq 0, \hspace{0.3cm} \text{or}  \hspace{0.3cm} \epsilon_{5} \leq \epsilon_{1} + \epsilon_{2} + \epsilon_{3}.
\end{equation}

\subsection{Error distributions in optical circuits vs superconducting qubits} \label{Error distributions}
If we perform Hardy's experiment in optical set-up \cite{Giuseppe1997,irvine2005,Lundeen2009,Yokota2009,vallone2011,chen2012,karimi2014,zhang2016,fan2017,ladder2017,Luo2018,Yang19}, errors can occur in many different ways, such as: (i) the preparation of the ideal quantum state, (ii)  in the construction of the measurement operators $A_{i}$ and $B_{j}$ where $i , j \in \left\lbrace 1,2\right\rbrace $ (as defined in Section \ref{Hardy_test}), (iii) due to the detection problems of the particles (this includes particle loss), etc. Introducing an error in Equation~(\ref{eq:1}) due to the above-mentioned reasons has a direct impact on Equation~(\ref{eq:4}), i.e., the logic of Hardy's argument ceases to work. Similarly, when Equation~(\ref{eq:2}) and Equation~(\ref{eq:3}) are non-zero, they make a contribution to the right-hand side of Equation~(\ref{eq:4}).  So, in an optical set-up, if Equation~(\ref{CHSH_noise_2}) is violated, it leads to the violation of local realism. No estimation of $q$ is required. The work \cite{irvine2005} does exactly this check of Equation~(\ref{CHSH_noise_2}) in its optical circuits.

But in the case of superconducting qubits, there are three types of errors, namely gate error, readout error, and multi-qubit gate error \cite{IBM}. Unlike in optical circuits, these errors are common to all superconducting qubits circuit and not specific to the circuits to test Hardy's paradox.
So, having an error in Equation~(\ref{eq:1}) does not have any impact on the right-hand side of Equation~(\ref{eq:4}) and Hardy's argument still works with this error. Similar reason applies when Equation~(\ref{eq:2}) and Equation~(\ref{eq:3}) are not zero. So, in this case, to test the violation of local realism, we will estimate the value of $q$ in Equation~(\ref{eq:8}) by a new method which will be described in the next section.

\subsection{Our proposed model for verifying whether $q > 0$ in superconducting qubits} \label{Estimate q}
We recall from Equation~(\ref{eq:8}) that $\epsilon_{5}=\epsilon_{4}+ q$. While doing the experiment, we can observe only the values of  $\epsilon_{5}$. 
 To estimate the value of $q$, we need to get the values of $\epsilon_{4}$. Now, the values of $\epsilon_{4}$ can be observed from the experiment directly for every MES as well as PS of two qubits as $q=0$, but it cannot be observed directly for NMES as $q>0$.
 
However, because of the nature of errors in superconducting qubits as explained in Section \ref{Error distributions}, the values of $\epsilon_{4}$ in both the cases (for $q=0$ \& $q > 0$) will follow the same distribution. So, its maximum value, say $\Sigma_{4}$, can be estimated from a large number of known MES and PS (with $q=0$), and then this estimate can be used in Equation~(\ref{eq:8}) to infer about $q$ for any unknown state as follows:
\begin{equation} \label{q_bound}
\epsilon_{5} \leq \Sigma_{4} + q, \hspace{0.3cm}  \text{or} \hspace{0.3cm}  q \geq \epsilon_{5} - \Sigma_{4}.
\end{equation} 
From Equation~(\ref{q_bound}), we can define the lower bound on $q$ as
\begin{equation} \label{q_lb}
q_{lb}=\epsilon_{5} - \Sigma_{4}. 
\end{equation}
 If  $q_{lb} > 0$, then Equation~(\ref{q_bound}) implies that Hardy's probability  $q > 0$.   

\subsection{Bounds in errors in optical circuits vs superconducting qubits}

For Hardy's test in optical circuits, if $\epsilon_{1}=\epsilon_{2}=\epsilon_{3}=\epsilon$, then from Equation~(\ref{CHSH_noise_2}), we can get the bounds in errors, i.e., $0 \leq \epsilon < \frac{1}{3}$ \cite{rabelo12}. But if $\epsilon_{1} \neq \epsilon_{2}\neq\epsilon_{3}$, then these bounds are not valid because there may be a case where any of $\epsilon_{1},\epsilon_{2},\epsilon_{3}$ can be close to one and the rest close to zero. Then theoretically the bounds in errors are between zero and one.

For Hardy's test in superconducting qubits, as explained in Section \ref{Error distributions}, all the error parameters $\epsilon_{1},\epsilon_{2},\epsilon_{3},\epsilon_{4},\epsilon_{5}$ follow the same distribution. So, the bounds in $\epsilon_{1},\epsilon_{2},\epsilon_{3},\epsilon_{4},\epsilon_{5}$ will also be the same, i.e., in between zero and one. But to verify Hardy's circuit, the maximum value of $\epsilon_{4}$, i.e., $\Sigma_{4}$ needs to be bounded. From Equation~(\ref{q_bound}), the theoretical bound of $\Sigma_{4}$ is $\left(1-q_{max}\right)=\left(1 - 0.09017\right) \approx 0.90983$. So, this bound is also valid for $\epsilon_{1},\epsilon_{2},\epsilon_{3}$. In the case of optical circuits, when $\epsilon_{1} \neq \epsilon_{2}\neq\epsilon_{3}$, the bounds are trivial, but for the case of superconducting qubits, the error bounds are defined independently, i.e., whether all of them takes the same value or not. It may be noted that if we get a value of any of $\epsilon_{1},\epsilon_{2},\epsilon_{3}$ greater than $0.90983$, then that circuit cannot be used for Hardy's test using superconducting qubits. But in optical circuits, in theory, if we get one of the values of $\epsilon_{1},\epsilon_{2},\epsilon_{3}$  greater than $0.90983$ and the rest zero, we still can perform Hardy's test.

The basic motivation for performing Hardy's test in superconducting qubits is to show the violation of local realism in superconducting qubits. Though Hardy's paradox is already tested in optical circuits~\cite{Giuseppe1997,irvine2005,Lundeen2009,Yokota2009,fedrizzi2011,vallone2011,chen2012,karimi2014,zhang2016,fan2017,ladder2017,Luo2018,Yang19}, none of them have been able to estimate the lower bound on Hardy's probability, i.e., $q_{lb}$ from their experiments. The advantage of performing Hardy's test in superconducting qubits is that we can estimate the lower bound on of Hardy's probability $q_{lb}$ as discussed in Section~\ref{Estimate q}.

\section{Circuits for Hardy's test using superconducting qubits} \label{Circuits for Hardy's equations}
We perform a series of experiments to check Hardy's non-locality for two qubits in the IBM quantum computer~\cite{IBM}. For simplicity, we use the \textit{ibmqx4}\cite{ibmqx4} chip which is five-qubit, as we only need two qubits for our experiment. This experiment can also be done using other chips consisting of any other number of qubits (more than or equal to two).  It uses a particular physical type of qubit called a superconducting transmon qubit made from superconducting materials niobium and aluminum, patterned on a silicon substrate. During all the experiments, the fridge temperature is maintained at $0.021$ K. Any experiment in the IBM quantum computer can be performed for 1 shot, 1024 shots, 4096 shots or 8192 shots in every run.

In the current IBM \textit{ibmqx4} chip topology~\cite{IBM}, for using multi-qubit gates like $CNOT$, there is a restriction, i.e., not all pairs of qubits can be used for circuit implementation. The list of possible combinations are given in details in the IBM website~\cite{IBM} and it is also discussed in Section \ref{Other_CNOT_Gate} in details. It should be noted that all the qubits are subject to different types of errors as given in the IBM website~\cite{IBM}. Initially, we  implement our circuit by choosing any possible pair of qubits and then validate the results for the rest of the possible combinations of qubits.

In~\cite{barun08}, a circuit consisting of two coupled electronic Mach-Zehnder (MZ) interferometers has been proposed for Hardy's test which is similar to the Hardy's original thought experiment~\cite{hardy92}. We implement this circuit in the IBM quantum computer. As described in~\cite{barun08}, there are three important parameters of this experiment: beam splitters $U_{B}(\theta)=\left(\begin{smallmatrix}$cos$\theta&-$sin$\theta\\ $sin$\theta& $cos$\theta\end{smallmatrix}\right)$, phase shifter $ U_{P}(\phi)=\left(\begin{smallmatrix}1&0\\0&e^{i\phi}\end{smallmatrix}\right)$, and the coupling $U_{C}(\phi)=\left(\begin{smallmatrix}1& & & \\ &1& & \\ & &1& \\ & & &e^{i2\phi}\end{smallmatrix}\right)$ which can be expressed as

\begin{gather}
\begin{aligned}
U_{B}(\theta)=&U_{3}(2\theta,0,0), \hspace{0.3cm}
U_{P}(\phi)=U_{1}(\lambda),\\
U_{C}(\phi)=&M_{3} \cdot CNOT \cdot M_{2} \cdot CNOT \cdot M_{1},
\end{aligned} \label{eq:12}
\end{gather}

where
\begin{gather*}
\begin{aligned}
U_{1}(\lambda)=&\left(\begin{matrix}1&0\\0&e^{\lambda i}\end{matrix}\right), \hspace{0.3cm}
U_{3}(\theta,\lambda,\phi)= \left(\begin{matrix} \text{cos} \frac{\theta}{2} &-e^{-\lambda i} \text{sin} \frac{\theta}{2}\\ e^{-\lambda i} \text{sin} \frac{\phi}{2} & e^{i\left( \phi +\lambda \right) }$ cos$\frac{\theta}{2}\end{matrix}\right),\\
M_{1}=&Id \otimes U_{1}(-\lambda), \hspace{0.3cm} M_{2}=U_{1}(\lambda) \otimes U_{1}(-\lambda), \hspace{0.3cm} M_{3}=Id \otimes U_{1}(2\lambda).
\end{aligned}
\end{gather*}

\noindent $CNOT$ is a controlled-NOT gate and $ Id $ is the identity gate. Here $U_{1}(\lambda) $, $U_{3}(\theta,\lambda,\phi)$, $CNOT$, and $Id$ are available as standard gates provided by the IBM quantum computer~\cite{IBM}. We decompose the coupling $U_{C}(\phi)$ by the standard IBM gates is shown in Equation~(\ref{eq:12}). 

\begin{figure}[h!]
\centering
\includegraphics[width=0.6\textwidth]{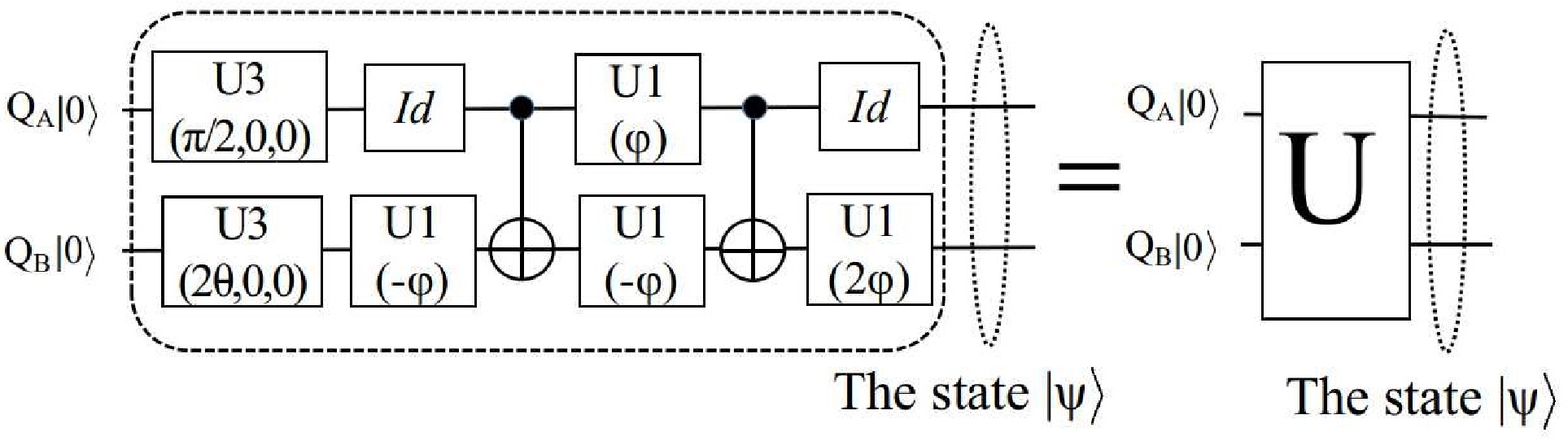}
\caption{The state $|\psi\rangle$ for Equation~(\ref{eq:13}).}
\label{fig:1}
\end{figure}

 For Hardy's test, the state  $\ket{\psi} $ for Alice and Bob is considered in~\cite{barun08} as
\begin{equation}
|\psi\rangle = V_{0}(V_{1} \otimes V_{2} )|00\rangle = \dfrac{\text{cos}\theta}{\sqrt{2}}(|00\rangle) + |10\rangle)+ \dfrac{\text{sin}\theta}{\sqrt{2}}(|01\rangle) + e^{i2\phi}|11\rangle), \label{eq:13}
\end{equation}
where  $V_{0}= U_{C}(\phi)$, $V_{1}=U_{B}\left( \dfrac{\pi}{4}\right)$, and $V_{2}=U_{B}(\theta)$.  The state $\ket{\psi}$ is expressed by the IBM gates is shown in Figure~\ref{fig:1} where $Q_{A}$ and $Q_{B}$ are the qubits for Alice and Bob respectively. The values of $\theta$ and $\phi$ in between $0$ to $90$ degrees for which $|\psi\rangle$ is found to be MES as well as PS are given in Table~\ref{table:1}. The measurements for Alice and Bob are described as follows. 

\begin{table}[h!]
\centering
\caption{MES and PS based on different values of $ \theta $ and $ \phi $ in between $0$ to $90$ degrees.} \label{table:1}
\begin{tabular}{ p{1.5cm}   p{1.5cm}  p{0.8cm}  p{5cm}    }
 \hline
   $\theta$ & $\phi$ & state & $|\psi\rangle $\\
 \hline 
 0  & any value & PS &  $\dfrac{1}{\sqrt{2}} (|0\rangle + |1\rangle) \otimes |0\rangle $ \\

 any value  & 0 & PS & $\dfrac{1}{\sqrt{2}} (|0\rangle + |1\rangle) \otimes $(cos$\theta |0\rangle + $sin$ \theta |1\rangle ) $\\

 90  & any value & PS & $\dfrac{1}{\sqrt{2}} (|0\rangle + e^{i2 \phi }|1\rangle) \otimes |1\rangle $\\

 45 & 90 & MES& $\dfrac{1}{2}(|00\rangle + |01\rangle + |10\rangle - |11\rangle)$ \\
\hline
\end{tabular}
\end{table} 

\begin{figure}[h!]
\centering
\includegraphics[width=0.35\textwidth]{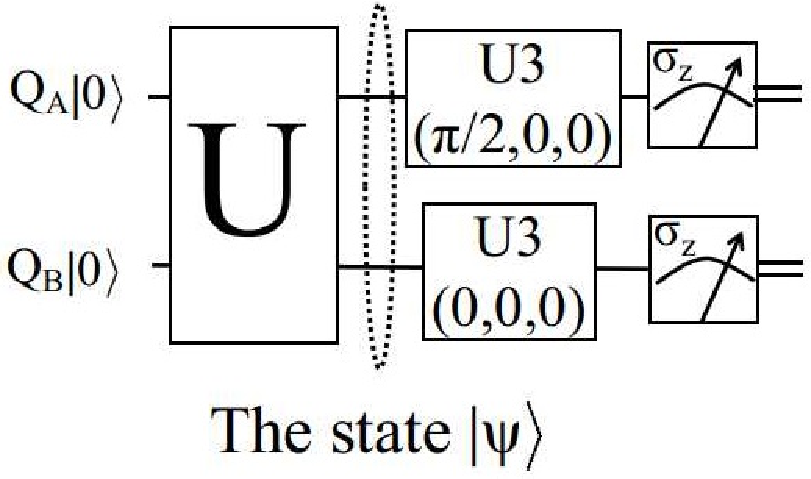}
\caption{Quantum circuit and measurement for $P(+1,+1|A_{1},B_{1})$ for Equation~(\ref{eq:5}).}
\label{fig:2}
\end{figure}

\begin{figure}[h!]
\centering
\includegraphics[width=0.45\textwidth]{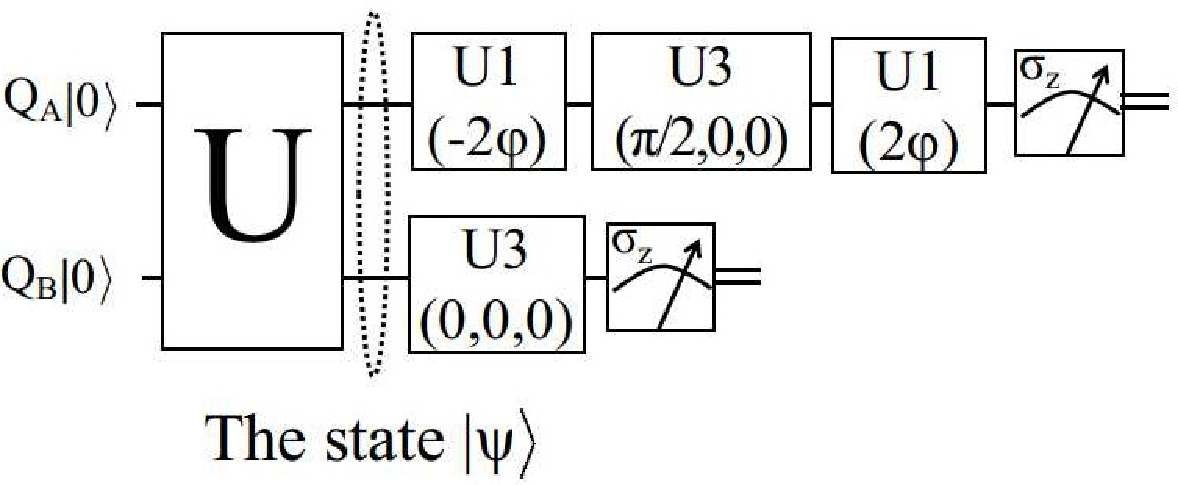} 
\caption{Quantum circuit and measurement for $P(+1,-1|A_{2},B_{1})$ for Equation~(\ref{eq:6}).}
\label{fig:3}
\end{figure}

\begin{figure}[h!]
\centering
\includegraphics[width=0.45\textwidth]{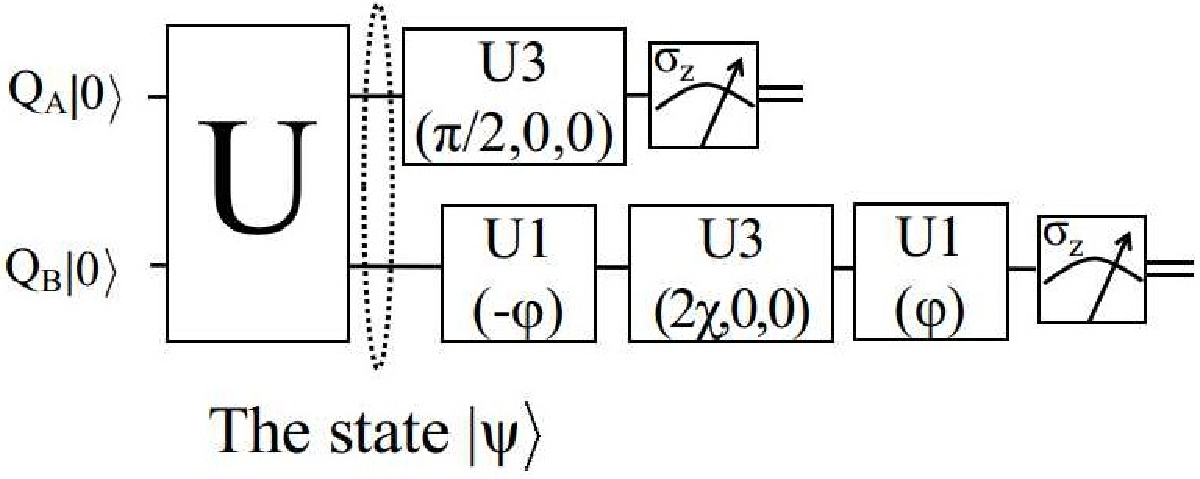} 
\caption{Quantum circuit and measurement for $P(-1,+1|A_{1},B_{2})$ for Equation~(\ref{eq:7}).}
\label{fig:4}
\end{figure}

\begin{figure}[h!]
\centering
\includegraphics[width=0.45\textwidth]{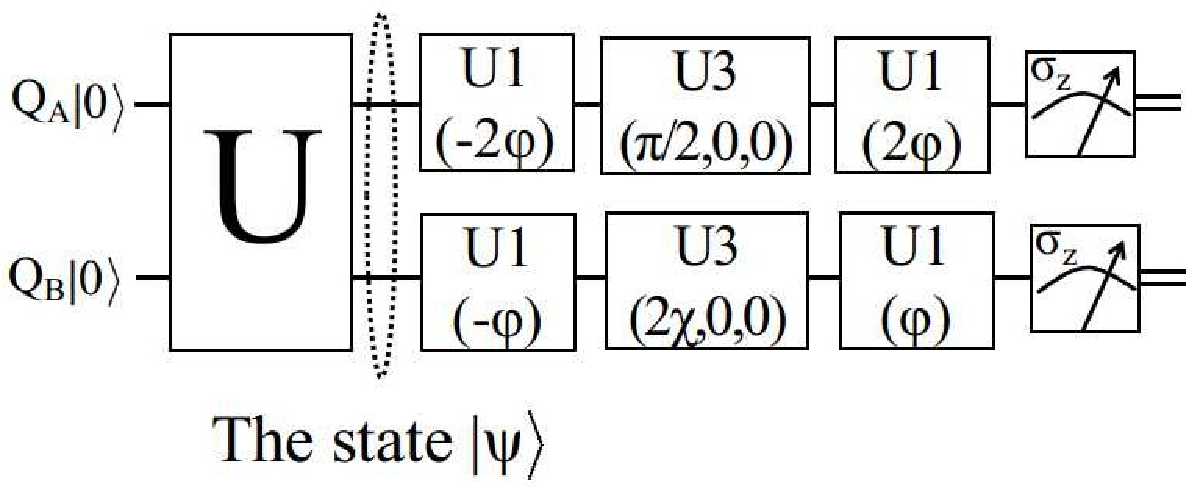} 
\caption{Quantum circuit and measurement for $P(+1,+1|A_{2},B_{2})$ for Equation~(\ref{eq:8}).}
\label{fig:5}
\end{figure}
\begin{figure}[t!]
\centering
\includegraphics[width=1.0\textwidth]{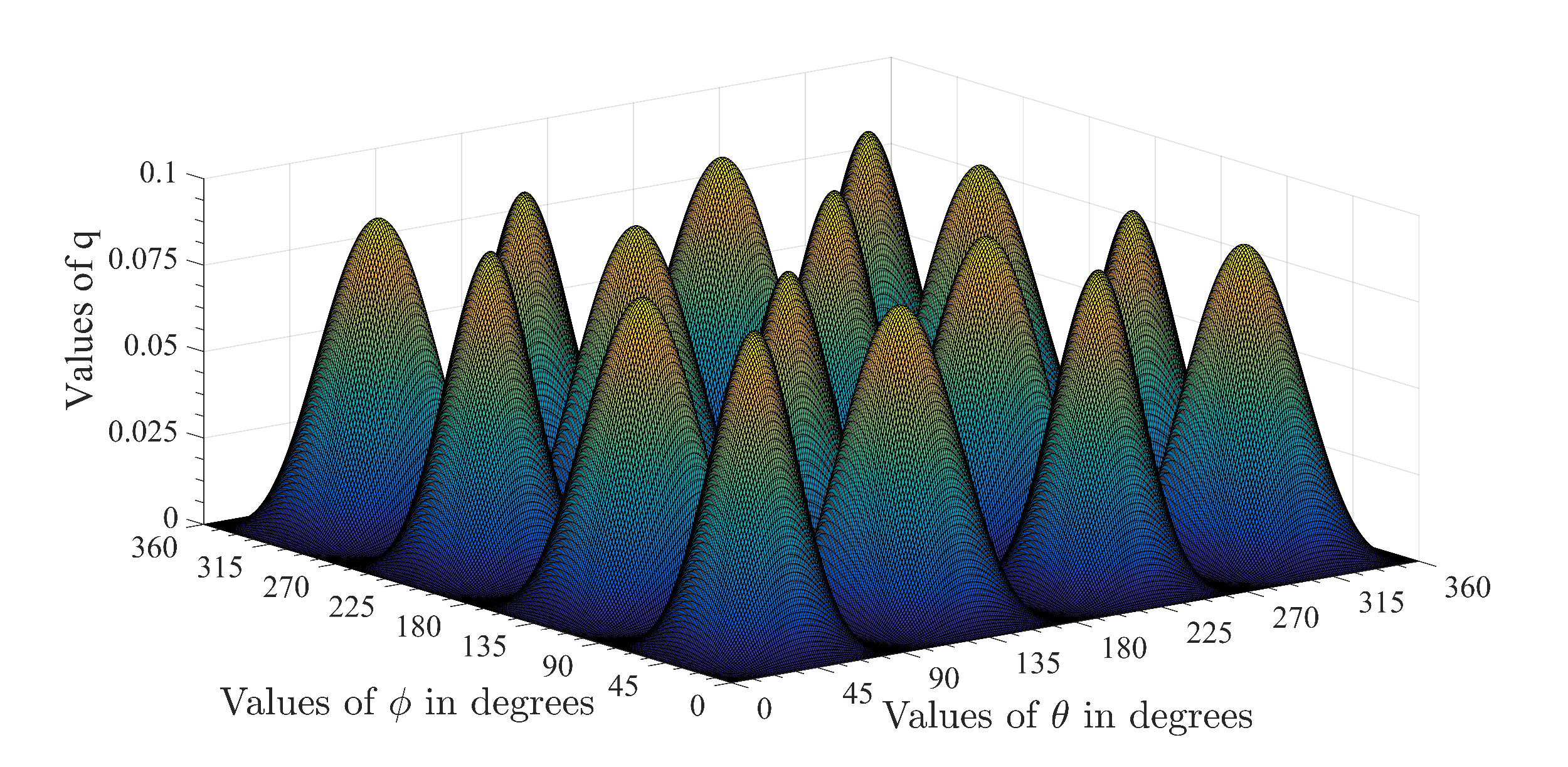}
\caption{The variation of $\theta$ and $\phi$ in degrees vs the values of $q$ from Equation~(\ref{eq:15}). }
\label{fig:6}
\end{figure}


\begin{gather} \label{measurements}
\begin{aligned}
A_{1}=& U_{B}\left( \dfrac{\pi}{4}\right) =U_{3}\left( \dfrac{\pi}{2},0,0\right) ,\\
B_{1}=& U_{B}\left( 0\right) =U_{3}\left( 0,0,0\right), \\
A_{2}=& U_{P}\left( 2\phi\right)  U_{B}\left( \dfrac{\pi}{4}\right)  U_{P}\left( -2\phi\right)
    =U_{1}(2\lambda) U_{3}\left( \dfrac{\pi}{2},0,0\right)  U_{1}(-2\lambda),\\
B_{2}=& U_{P}(\phi) U_{B}(\chi) U_{P}(-\phi)
    = U_{1}(\lambda) U_{3}\left( 2\chi,0,0\right) U_{1}(-\lambda), 
\end{aligned}
\end{gather}
where $\text{cot}\chi = \text{tan}\theta\text{cos}\phi$.  The measurements are done in $\sigma_{z}$ basis. The experimental circuits of Equations~(\ref{eq:1}), (\ref{eq:2}), (\ref{eq:3}), and~(\ref{eq:4}) for the state $\ket{\psi}$ using the above measurements in the IBM quantum computer are given in Figures~\ref{fig:2},~\ref{fig:3},~\ref{fig:4}, and~\ref{fig:5} respectively. The theoretical value of $P(+1,+1|A_{2},B_{2})$  is given by $\left|\braket{\psi|A_{2} \otimes B_{2} | \psi} \right| ^{2}$, which using Equation~(\ref{eq:13}) and Equation~(\ref{measurements}) becomes
\begin{equation}
q=\left|\frac{1}{2}\text{cos} \theta \text{cos} \chi \left( 1-e^{-2i\phi}\right) \right| ^{2}.
\label{eq:15}
\end{equation}
The maximum value of $q$, i.e., $q_{max}$ is found from Equation~(\ref{eq:15}) when
\begin{equation}
\text{cos}(2\theta)=\text{cos}(2\phi)=2-\sqrt{5}. \label{eq:14}
\end{equation}
 If the variation of $\theta$ and $\phi$ are carried out in between $0$ to $90$ degrees, then $q_{max}$ occurs at $\theta=\phi=51.827$ degrees approximately. But in general similar analysis can be done for any values of $\theta$ and $\phi$. In Figure~\ref{fig:6}, we plot~\cite{Matlab} the values of $q$  from Equation~(\ref{eq:15}) by varying $\theta$ and $\phi$ from $0$ to $360$ degrees. In this figure, we can see that $q_{max}$ is achieved for $\theta=\phi=51.827$ degrees and also for other values of $\theta$ and $\phi$ such that Equation~(\ref{eq:14}) is satisfied. 

 For $\phi=90$ and $\theta \neq \left\lbrace 0,45,90 \right\rbrace $ degrees, from Equation~(\ref{eq:13}), we get $|\psi\rangle$ as NMES. This means, if we perform Hardy's test, a non-zero value of $q$ in Equation~(\ref{eq:4}) has to be found. But when $\phi=90$ degree, we get $\chi=90$ degree which means the right-hand side of Equation~(\ref{eq:15}) is zero. So, in this experimental set-up, Hardy's test fails for all NMES for the values of $\phi=90$ and $\theta \neq \left\lbrace 0,45,90 \right\rbrace $ degrees.
  
\section{How should we perform statistical analysis of experimental results?} 
It is known from the central limit theorem~\cite{feller}  that if $\lbrace X_{1},X_{2},\ldots, X_{n}\rbrace $ are independent and identically distributed random samples drawn from any population with mean $\mu$ and variance $\sigma^{2}$ and if $n$ is large, then the sample mean $ \bar{X}=\frac{1}{n} \sum^{n}_{i} X_{i}$ follows a normal distribution with mean $\mu$ and variance $\sigma^{2}/n$, i.e., $ \bar{X} \backsim N (\mu, \sigma^{2}/n) $. It follows that the variable
\begin{equation*}
Z=\dfrac{\bar{X}-\mu}{\sqrt{\sigma^{2}/n}}, 
\end{equation*}
follows the standard normal distribution, i.e., $Z \backsim N(0,1)$.
If the population variance $\sigma^{2}$ is unknown, it is replaced by its closest approximation, i.e., the sample variance $S^{2}$. Then the quantity follows a Students t-distribution~\cite{feller} described as
\begin{equation*}
T=\dfrac{\bar{X}-\mu}{\sqrt{S^{2}/n}}, \hspace{0.3cm} \text{where} \hspace{0.3cm} S^2=\dfrac{1}{n-1} \sum_{i=1}^{n}(X_{i}-\bar{X})^{2}.
\end{equation*} 
Moreover, while normal approximation works only for very large $n$ (ideally infinite), the Students t-distribution holds for small $n$ as well. This distribution is a function of the degrees of freedom, which is one less than the number of times the experiment is repeated. As the number of degrees of freedom tends to infinite, the Student's t-distribution converges to the normal distribution.

\subsection{Hypothesis testing and confidence interval}
We know that it is impossible to conduct any experiment without any error. So, when the conclusion is drawn from an experimental result, it is not appropriate to claim that the results are  $100\%$ correct. Statistically speaking, we can only test a hypothesis and conclude the test based on our experimental results with some percentage of confidence in our conclusion.
In hypothesis testing~\cite{feller}, we turn a question of interest into a hypothesis about the value of a parameter or a set of parameters. In our case, suppose we want to test whether a given state is NMES or not. Then, according to the notion of hypothesis testing, we can have the following formulation from Equation~(\ref{q_bound}).

Test the null hypothesis 
$$\mathcal{H}_{0}: q = 0  \hspace{0.5cm} \text{(The unknown state is not an NMES)}$$
against the alternative hypothesis
$$\mathcal{H}_{1}: q > 0  \hspace{0.5cm} \text{(The unknown state is an NMES)}.$$

So it is evident that our test will be a one-sided test~\cite{onesided} as $q$ is a non-negative quantity. The significance level $\alpha$ is equal to the false-positive error or Type I error which is defined by the probability $\Pr$(reject $\mathcal{H}_{0} \mid \mathcal{H}_{0}$ is correct) where $ 0 \leq \alpha \leq 1 $. The level of confidence is defined by $\left( 1-\alpha\right)$ or $100(1-\alpha)\%$. 
Let us assume that an experiment is repeated $n$ number of times, where $\mu$, $ \bar{X}$, $\sigma$,  and $S$  are the population mean, sample mean, population standard deviation and sample standard deviation respectively. 
Again, let $100(1-\alpha)\%$  confidence interval (CI) of $\bar{X}$  be the interval $\left[  X_{lb} ,  X_{ub} \right] $. 
This means that we have $100(1-\alpha)\%$ confidence that $\mu$ will lie in between $ X_{lb}$ and $ X_{ub}$.

For the standard normal distribution, the expression for $100(1-\alpha)\%$ CI for the mean in the above situation is $\left(  \bar{X} \pm z_{\frac{\alpha}{2}}\frac{\sigma}{\sqrt{n}} \right) $, where $z_{\frac{\alpha}{2}}$ is the value of the standard normal variable $Z$ such that
\begin{equation*}
\text{Pr}(-z_{\frac{\alpha}{2}} < Z < z_{\frac{\alpha}{2}})=(1-\alpha).
\end{equation*}
For this case, $ X_{lb} =\left(  \bar{X} - z_{\frac{\alpha}{2}}\frac{\sigma}{\sqrt{n}} \right)  $ and  $ X_{ub} =\left(  \bar{X} + z_{\frac{\alpha}{2}}\frac{\sigma}{\sqrt{n}} \right) $. Here, $z_{\frac{\alpha}{2}}$ depends only on the value of $\alpha$.
For the Student's t-distribution, the expression for $100(1-\alpha)\%$ 	 CI is given by $\bar{X} \pm t_{\frac{\alpha}{2}}\frac{S}{\sqrt{n}} $ where $t_{\frac{\alpha}{2}}$ is given by the following expression
\begin{equation*}
\text{Pr}(-t_{\frac{\alpha}{2}} < T< t_{\frac{\alpha}{2}})=(1-\alpha).
\end{equation*}
For this case, $ X_{lb} =\left(  \bar{X} - t_{\frac{\alpha}{2}}\frac{S}{\sqrt{n}} \right)  $ and  $ X_{ub} =\left(  \bar{X} + t_{\frac{\alpha}{2}}\frac{S}{\sqrt{n}} \right) $. Here, $t_{\frac{\alpha}{2}}$ depends on the value of $\alpha$ and the degrees of freedom $\nu = \left( n -1\right)$, where $n$ is the number of samples used for the experiment. As $\nu \to\infty $, we have $t_{\frac{\alpha}{2}} \to  z_{\frac{\alpha}{2}}$.  
\subsection{How to calculate the difference between two Student's t-distributed variables?} \label{student's t calculation}
To calculate $q_{lb}$ from Equation~(\ref{q_lb}), we need to calculate the statistics for the difference of two random variables $X$ and $Y$ corresponding to the experimental values of $\epsilon_{5}$ and $\Sigma_{4}$ respectively.
Here both $X$ and $Y$ are assumed to follow standard normal distributions. Also, for $n$ number of samples, let the sample means of $X$ and $Y$ be $\bar{X}$ and $\bar{Y}$ respectively and the sample standard deviations be $S_{X}$ and $S_{Y}$ respectively. Then for some value of $\alpha$, let $ \left( \bar{X} \pm t_{X} \right) $ where $t_{X}=t_{\frac{\alpha}{2}}\frac{S_{X}}{\sqrt{n}} $ represents the $(1-\alpha)\%$ CI  around the mean of $X$. Similarly, let $ \left( \bar{Y} \pm t_{Y} \right) $  where $t_{Y}=t_{\frac{\alpha}{2}}\frac{S_{Y}}{\sqrt{n}} $ represents the same for the random variable $Y$. Now if we want to calculate the value of another Student's t-distributed random variable $W$ such that $W=X-Y$, then the sample mean $\bar{W}$ of $W$ is given by $\bar{W}=\left( \bar{X}-\bar{Y}\right)$ and the sample standard deviation is given by $S_{W}=\sqrt{S_{X}^{2}+S_{Y}^{2}}$. Then the $(1-\alpha)\%$ CI around mean of $W$ is $ \left( \bar{W} \pm t_{W} \right) $  where $t_{W}=t_{\frac{\alpha}{2}}\frac{S_{W}}{\sqrt{n}} $. Now the value of $W_{lb}$ will be 
\begin{equation} \label{zmin}
W_{lb}=\left( \bar{W}-t_{W}\right) = \left( \bar{X}-\bar{Y}-t_{\frac{\alpha}{2}}\frac{\sqrt{S_{X}^{2}+S_{Y}^{2}}}{\sqrt{n}}\right).
\end{equation}
We can use Equation~(\ref{zmin}) to estimate $\hat{q}_{lb}$ of $q_{lb}$ as defined in Equation~(\ref{q_lb}) as
\begin{equation} \label{hatqlb}
\hat{q}_{lb} = \left( \bar{\epsilon}_{5}-\bar{\Sigma}_{4}-\Delta\right) \hspace{0.3cm} \text{where} \hspace{0.3cm} \Delta=t_{\frac{\alpha}{2}}\frac{\sqrt{S_{\epsilon_{5}}^{2}+S_{\Sigma_{4}}^{2}}}{\sqrt{n}}.
\end{equation}
Here $\bar{\epsilon}_{5}$ and $S_{\epsilon_{5}}$ are the sample mean and sample standard deviation of $\epsilon_{5}$ respectively over $n$ number of samples. Similarly, $\bar{\Sigma}_{4}$ and $S_{\Sigma_{4}}$ are the corresponding quantities for $\Sigma_{4}$.
When the value of $\alpha$ increases, the confidence in data $(1-\alpha)$ decreases. So, the value of $t_{W}$ increases because the value of $t_{\frac{\alpha}{2}}$ is computed from the minimum side of the distribution and as per Equation~(\ref{zmin}), the value of $W_{lb}$ decreases. Thus, in Equation~(\ref{hatqlb}), as $\alpha$ increases, $\Delta$ increases, and so $\hat{q}_{lb}$ decreases. Note that, all the statistical analysis is done for certain fixed data~\cite{feller}.

Given an unknown state, to find whether that state is NMES or not, create an experimental set-up where Equations~(\ref{eq:1})-(\ref{eq:3}) are satisfied theoretically and Equations~(\ref{eq:5})-(\ref{eq:7}) are validated experimentally. Then we use a two-phase procedure as presented in Algorithm~\ref{alg1}. 
\RestyleAlgo{boxed}
\begin{algorithm}[h!] 
\textbf{OFF-LINE PHASE (Estimation of $\bar{\Sigma}_{4}$ from unknown MES \& PS)}: \\
\nl Do the experiment for Equation~(\ref{eq:8}) for $k$ numbers of known MES as well as PS to get the values of $\bar{\epsilon}_{4}$.\\
\nl Then from that data, calculate the values of $\bar{\Sigma}_{4}= \max \lbrace\bar{\epsilon}_{4}\rbrace$ and $S_{\Sigma_{4}}$.\\ \ \\
\textbf{ON-LINE PHASE (Estimation of $\hat{q}_{lb}$ from the unknown state)}:\\ 
\nl  Do the experiment for Equation~(\ref{eq:8}) to calculate $\bar{\epsilon}_{5}$ and $S_{\epsilon_{5}}$ for the unknown state. \\
\nl Calculate the value of $\Delta$ from Equation~(\ref{hatqlb}). \\
\nl Plug-in the values of $\bar{\epsilon}_{5}$, $\bar{\Sigma}_{4}$ and $\Delta$ in Equation~(\ref{hatqlb}) to get an estimate $\hat{q}_{lb}$ of the lower bound $q_{lb}$ on $q$.\\
\nl \If{$\hat{q}_{lb} > 0$ (i.e., if $\bar{\epsilon}_{5} > \bar{\Sigma}_{4} + \Delta$ )} {
the unknown state is NMES and the value of Hardy's probability $q$ of that unknown state is greater than or equals to the value of $\hat{q}_{lb}$, i.e., $q \geq \hat{q}_{lb}$} 
\Else{
no decision can be made about the unknown state.}
\caption{Estimation for the lower bound $q_{lb}$ on $q$, i.e., $\hat{q}_{lb}$ for superconducting qubits.}
\label{alg1}
\end{algorithm}

In this way, we can identify whether an unknown state is NMES or not with some confidence. If it is NMES, then we can estimate a lower bound on Hardy's probability of that NMES.
The two-phase process is similar to the scenario of classical error-control coding~\cite{shanon47} wherein off-line phase channel noise is estimated using known messages and subsequently that estimation is used in the on-line phase to correct the transmission error of unknown messages. 

\section{Revisiting known experimental results and their statistics}
In this section, first we revisit important experimental works on Hardy's test in optical set-up. Next, we look into significant experiments performed in IBM quantum computer. 

\subsection{Prior works on Hardy's test in optical set-up}

In this section, we will survey the major experiments performed for Hardy's test using optical circuits so far. Their set-up and the results are summarized in Table~\ref{table:2}.

\begin{table} [h!]
\centering
\caption{Summary of prior works on Hardy's test in optical set-up where SD=standard deviation and CI=confidence interval.} \label{table:2}
\begin{tabular}{  p{3cm}| p{0.7cm}| p{2cm}| p{2cm}| p{1.7cm}| p{1.3cm}}
\hline
 Authors & Year  & Distribution of data  & number of samples mentioned? & Form: $\text{mean} \pm \text{SD}$   & CI Representation \\
\hline
 Irvine \textit{et al.} \cite{irvine2005} & 2005 & Not specified & Not specified & Yes & No\\ 
 \hline
  Lundeen \textit{et al.} \cite{Lundeen2009} & 2009 & Poissonian & Not specified & Yes & No\\ 
 \hline
  Yokota \textit{et al.} \cite{Yokota2009} & 2009 & Poissonian & Not specified & Yes & No\\ 
 \hline
  Fedrizzi \textit{et al.}  \cite{fedrizzi2011} & 2011& Not specified & Not specified &Yes & No\\ 
 \hline
  Vallone \textit{et al.}  \cite{vallone2011} & 2011& Not specified & Not specified & Yes& No\\ 
 \hline
  Chen \textit{et al.} \cite{chen2012} & 2012 & Not specified  & Not specified& Yes&No\\ 
 \hline
 karimi \textit{et al.} \cite{karimi2014} & 2014 & Poissonian &Not specified  & Yes&No\\ 
 \hline
 Zhang \textit{et al.}  \cite{zhang2016} &2016 & Not specified  & Not specified& Yes&No\\ 
 \hline
Fan  \textit{et al.} \cite{fan2017}  & 2017 & Poissonian & Not specified& Yes&No\\ 
 \hline
Chen \textit{et al.} \cite{ladder2017} & 2017 & Not specified  & Not specified& Yes&No\\ 
 \hline
 Luo \textit{et al.} \cite{Luo2018} & 2018 & Poissonian  & Not specified& Yes&No\\ 
 \hline
 Yang \textit{et al.} \cite{Yang19} & 2019 & Poissonian  & Not specified& Yes&No\\ 
 \hline
\end{tabular}
\end{table}

In \cite{irvine2005,fedrizzi2011,vallone2011,chen2012,zhang2016,ladder2017}, the authors have represented their coincidence count in mean $\pm$ standard deviation form but they did not specify the distribution of the sample data and the number of samples taken. They also did not represent the confidence on their data.

In \cite{Lundeen2009,Yokota2009,karimi2014,fan2017,Luo2018,Yang19}, the authors have mentioned that they have assumed that all the error bars follow the Poissonian statistics for the coincidence count and they have represented their data in mean $\pm$ standard deviation form. But again, they did not mention the distribution of samples and number of samples they have used and confidence on their data.

 \subsection{Significant experiments performed in superconducting qubits}

\begin{table}[h!]
\centering
 \caption{Significant experiments performed in IBM five-qubit quantum computer and their error statistics where $n$= number of times the experiment has been performed, $s$=number of shots in each time, SD=standard deviation, and CI=confidence interval.} \label{table:3}
\begin{tabular}{p{4cm}| p{0.7cm}| p{0.5cm}|p{1cm}| p{1.8cm}| p{1.8cm}}
\hline
 Authors & Year  & $n$ & $s$  & Form:  $\text{mean} \pm \text{SD}$  & CI Representation \\
\hline 
S. J. Devitt \cite{devitt16} & 2016 & 1 &8192 & No  & No\\ 
 \hline
 Alsina \textit{et al.}  \cite{mermin2016} & 2016  & 1 & 8192 & Yes & No\\ 
 \hline
 Berta \textit{et al.} \cite{berta2016} & 2016  & 1 & 8192 & No & No\\  
 \hline
   Huffman \textit{et al.} \cite{huffman2017} & 2017  &  10& 8192 &Yes &No\\ 
 \hline
  Hebenstreit  \textit{et al.} \cite{alsina2017} & 2017  & 1 & 8192 & No & No\\ 
 \hline
Behara \textit{et al.}  \cite{Qcheque} & 2017  & 1& 1024, 4096, 8192 & No&No\\ 
 \hline
   Yal\ifmmode \mbox{\c{c}}\else \c{c}\fi{}\ifmmode \imath \else \i \fi{}nkaya \textit{et al.} \cite{Yalcinkaya17} & 2017 & 5& 8192& Yes & No\\ 
 \hline
     W. Hu   \cite{hu2018} & 2018 &  1 & 8192 & No & No\\ 
 \hline
  Gangopadhyay \textit{et al.}  \cite{DJALGOPani18} & 2018 & 10 & 8192 & Yes & No\\
  \hline
  Lee \textit{et al.}  \cite{Lee19} & 2019 & 1 & 1024 & No & No\\
  \hline
\end{tabular}
\end{table} 
 We present a survey of the other experiments done in the IBM quantum computer and their statistical representation of the results which are summarized in Table~\ref{table:3}.
In \cite{devitt16,berta2016,mermin2016}, the authors have represented the error in their experiments by the standard formula  $\sqrt{p(1-p)/8192}$, where $p$ is the estimate of the probability of a given measurement outcome in a given experiment. However, some authors~\cite{mermin2016,huffman2017,Yalcinkaya17,DJALGOPani18} have represented their data in the form of mean $\pm$ standard deviation but none of them have represented their data in CI form.

\section{Our Experimental Results and Discussion}  \label{Experimental Results and Discussion}

The confidence interval around the mean for Student's t-distribution depends upon two quantities. 
\begin{enumerate}
\item The degrees of freedom for Student's t-distribution which is one less than the number of samples taken for the experiment.
\item The percentage of confidence we need on our data.
\end{enumerate} 
\noindent Based on these two quantities, we get the confidence interval around the mean for any data. 

In the IBM five-qubit quantum computer, any experiment can be performed for 1 shot or one of 1024, 4096 and 8192 (which is the maximum available) shots. It means that these many number of times the experiment is performed internally and the average of that is reported as output. But if we repeat any experiment with 8192 shots for a few times, we may see a significant deviation among each of the average values. Further, if we perform any experiment for 8192 shots, then the IBM quantum experience does not provide us with all the samples of those shots, rather it only provides the mean value. So if we perform any experiment only once with 8192 shots and make a conclusion based on the result, then that is not a statistically correct way. Instead, if someone does the experiment $n$ number of times with 8192 shots/time and then estimates the mean and the standard deviation assuming Student's t-distribution with required CI, then that would be statistically more accurate. We will show the variation of the same result by
 varying $n$ and the percentage of CI for Hardy's experiment.
 
 \begin{figure}[t!]
\centering
\includegraphics[width=10cm]{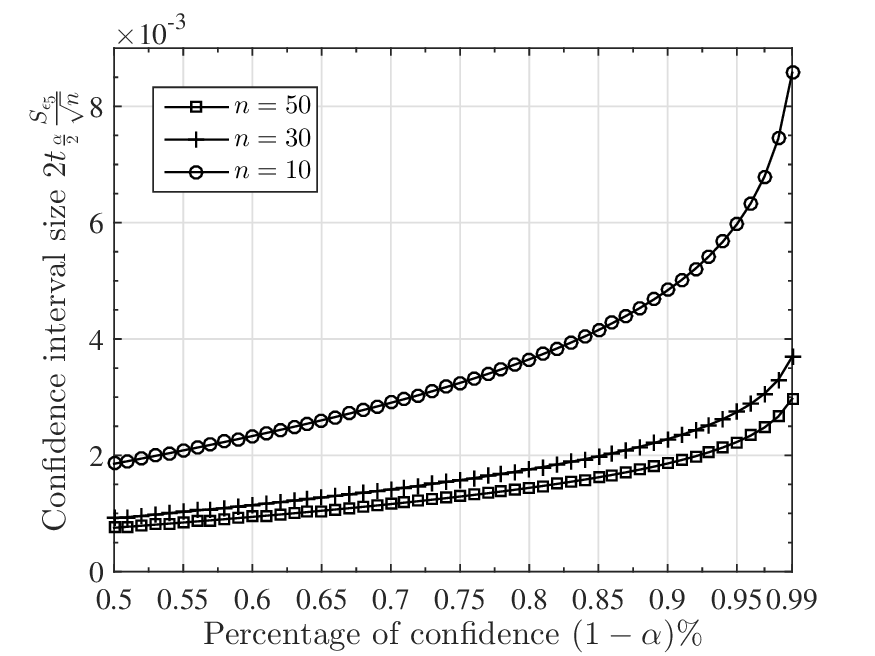}
\caption{The variation of the confidence interval size $\left( 2t_{\frac{\alpha}{2}}\frac{S_{\epsilon_{5}}}{\sqrt{n}} \right) $ with the percentage of confidence $\left( 1 - \alpha\right) \% $ with varying number of samples is $n \in \lbrace10,30,50\rbrace$.}
\label{fig:7}
\end{figure} 

In Figure~\ref{fig:7}, we show the variation of the confidence interval size $ 2t_{\frac{\alpha}{2}}\frac{S_{\epsilon_{5}}}{\sqrt{n}}  $ with the percentage of confidence $\left( 1 - \alpha\right) \% $ with varying number of samples $n \in \lbrace10,30,50\rbrace$. Each of the samples is the mean of Hardy's experiment run for 8192 shots. It can be seen that the confidence interval size increases with a decreasing number of samples. Also, for a fixed number of samples, the interval size increases with an increasing percentage of confidence in data. From this result, it can be concluded that if an experiment in the IBM quantum computer is done only on one sample for 8192 shots, then that would not be a statistically correct way of representation since each of the samples can deviate from the mean significantly. 
As we have limited access to the IBM quantum computer, we take $n=10$ and the experiments are run for MES, PS, and NMES for some values of $\theta$ and $\phi$ in between $0$ to $90$ degrees as shown in Table~\ref{table:4} and Table~\ref{table:5}. 

We perform experiments with different combinations of qubits for Alice and Bob and discuss the results below. 

\subsection{Experiments for Hardy's non-locality using qubit $Q_{3}$ for Alice and $Q_{4}$ for Bob}
In this section, we check Hardy's non-locality by taking a specific combination of qubits from all available combinations in the IBM quantum computer. Later we will discuss rest of the possible combinations.

\subsubsection{Experimental validation of the circuit for Hardy's test}
We perform the experiments for Equations~(\ref{eq:5}),~(\ref{eq:6}), and~(\ref{eq:7}) (Figures~\ref{fig:2},~\ref{fig:3} and~\ref{fig:4} respectively) for the values of $\theta$ and $\phi$ in degrees. The values of $\bar{\epsilon}_{1}$, $\bar{\epsilon}_{2}$ and $\bar{\epsilon}_{3}$ in each of the experiments are found to be less than $0.1$, implying that Equations~(\ref{eq:5}),~(\ref{eq:6}) and~(\ref{eq:7}) are satisfied. These results indicate that this experimental set-up is now valid for Hardy's test. As discussed in Section \ref{Practical Hardy's test}, for practical verification of Hardy's test using superconducting qubits, only one experiment for  Equation~(\ref{eq:8}) (Figure~\ref{fig:5}) has to be performed to establish non-locality, that is why the details of the results for $\bar{\epsilon}_{1}$, $\bar{\epsilon}_{2}$ and $\bar{\epsilon}_{3}$ are not presented. The experimental results of Equation~(\ref{eq:8}) are presented in Table~\ref{table:4} and Table~\ref{table:5} for some selected values of $\theta$ and $\phi$. 

\begin{table} [t!]
\centering 
\caption{Results of $\bar{\epsilon}_{4}$, standard deviations (SD) and the values of $  t_{\frac{\alpha}{2}}\frac{S_{\epsilon_{4}}}{\sqrt{n}}  $ for different confidence intervals (CIs) with $n=10$ for some values of $\theta$ and $\phi$ for which  MES as well as PS are created for the pair of qubits $\left( Q_{3}, Q_{4}\right) $.}
\begin{tabular}{ p{0.7cm}| p{0.9cm}| p{1.1cm}| p{0.9cm}|  p{1.2cm}| p{1.2cm}| p{1.2cm}| p{1.2cm}} 
\hline
 State & $\theta$, $\phi$  & $\bar{\epsilon}_{4}$ & SD & \multicolumn{4}{c}{ $t_{\frac{\alpha}{2}}\frac{S_{\epsilon_{4}}}{\sqrt{n}}$ for different CIs}\\
 \cline{5-8} & & ($q=0$) & ($S_{\epsilon_{4}}$)&  $99\%$ & $95\%$ & $90\%$ & $80\%$ \\
\hline 
 MES  & 45, 90 & 0.0807 & 0.0037 & 0.003802	& 0.002647 &	0.002145 &	0.001618\\
\hline 
PS & 0, 0  &0.0193 & 0.0014 & 0.001439 & 0.001001 & 0.000812 & 0.000612 \\
\hline
PS & 90, 0  &0.0209 & 0.0015 & 0.001542 &0.001073  & 0.00087 & 0.000656\\
\hline
PS & 45, 0  & 0.0217 & 0.0019 & 0.001953  & 0.001359  & 0.001101 & 0.000831\\
\hline
PS & 90, 45  & 0.0282 & 0.0013 & 0.001336 & 0.00093 & 0.000754 & 0.000569 \\
\hline
\end{tabular}
\label{table:4}
\end{table}

\begin{table}[t!]
\centering
 \caption{Results of $\bar{\epsilon}_{5}$, standard deviations (SD) and the values of $ t_{\frac{\alpha}{2}}\frac{S_{\epsilon_{5}}}{\sqrt{n}} $ for different confidence intervals (CIs) with $n=10$ for some values of $\theta$ and $\phi$ for which NMES are created for the pair of qubits $\left( Q_{3}, Q_{4}\right) $.} \label{table:5}
\begin{tabular}{ p{0.9cm}| p{0.9cm}| p{1.2cm}| p{0.9cm}| p{1.2cm}| p{1.2cm}| p{1.2cm}| p{1.2cm}}
\hline
 State & $\theta$, $\phi$  & $\bar{\epsilon}_{5}$ & SD & \multicolumn{4}{c}{ $t_{\frac{\alpha}{2}}\frac{S_{\epsilon_{5}}}{\sqrt{n}}$ for different CIs}\\
 \cline{5-8} & & $(q > 0)$ & ($S_{\epsilon_{5}}$)&  $99\%$ & $95\%$ & $90\%$ & $80\%$ \\
\hline 
NMES & 51.827, 51.827  &0.1281 & 0.0039 & 0.004008 & 0.00279  &  0.002261 & 0.001706 \\
\hline
 NMES & 55, 55  &0.1273 &  0.0045 & 0.004625 &  0.003219 & 0.002609 & 0.001968  \\
  \hline
NMES & 45, 45  &0.1041 &  0.0044 & 0.004522 & 0.003148 &  0.002551 &  0.001924 \\
 \hline 
 NMES & 30, 60   &0.0832 & 0.0052 &  0.005344 &  0.00372 & 0.003014 & 0.002274 \\
 \hline
   NMES & 60, 30  &0.0553 & 0.0028 & 0.002878 & 0.00200 & 0.001623 & 0.001225\\
\hline
NMES & 10, 80 & 0.067 & 0.0038 & 0.00391 & 0.00272 & 0.0022 & 0.00166 \\ 
 \hline
  NMES & 80, 10 &0.0241 & 0.0016 & 0.001644 & 0.001145 & 0.000927 &  0.0007\\
 \hline
\end{tabular}
\end{table}

\begin{table}[h!]
\centering
 \caption{Comparison of the results of $q$ and $\hat{q}_{lb}$ for some values of $\theta$ and $\phi$ for which NMES is created for different confidence interval (CIs) for the pair of qubits $\left( Q_{3}, Q_{4}\right) $}
\begin{tabular}{ p{0.9cm}| p{0.9cm}| p{1cm}|  p{1.5cm}| p{1.5cm}| p{1.5cm}| p{1.5cm}}
\hline
 State & $\theta$, $\phi$  & $q$  & \multicolumn{4}{c}{ $\hat{q}_{lb}$ for different CIs }\\
 \cline{4-7} & & &  $99\%$ & $95\%$ & $90\%$ & $80\%$  \\
\hline
NMES & 51.827, 51.827  & 0.09017  & 0.041876 & 0.043554 & 0.044283 & 0.045049 \\
\hline
 NMES & 55, 55  & 0.0886  & 0.040613 & 0.042432 & 0.043222 & 0.044052 \\
 \hline 
NMES & 45, 45  & 0.0833  & 0.017492 & 0.019287 &  0.020067 & 0.020886 \\
 \hline 
 NMES & 30, 60   &0.0433  & -0.004058 & -0.002066 & -0.001199 & -0.000290\\
 \hline
   NMES & 60, 30  &0.0433  & -0.030168 & -0.028718 & -0.02809 & -0.027429 \\
\hline
NMES & 10, 80 &  0.00088 & -0.019154 & -0.017495 & -0.016773 &  -0.016018 \\ 
 \hline
  NMES & 80, 10 & 0.00088 & -0.060742 & -0.059484 & -0.058937 &  -0.058363\\
 \hline
\end{tabular}
\label{table:6}
\end{table}

\subsubsection{Test of non-locality when $q=q_{max}$} \label{TestofNL_q=qmax}
From Table~\ref{table:4}, for the MES, i.e., $\theta=45$ and $\phi=90$ degrees, we get $\bar{\epsilon}_{4}=0.0807$, $S_{\epsilon_{4}}=0.0037$, and the values of $t_{\frac{\alpha}{2}}\frac{S_{\epsilon_{4}}}{\sqrt{n}}$ for different CIs. We take four possible CIs: $99\%$, $95\%$, $90\%$, and $80\%$. Similarly, for the PS, we get $\bar{\epsilon}_{4}$ to be less than $0.03$ (there are only one MES possible but ideally an infinite number of PS possible as shown in Table~\ref{table:1}. But due to limited access to the IBM quantum computer, we take $k=50$ number of PS as indicated in Algorithm 1. As for all the PS, we get $\bar{\epsilon}_{4}$ to be less than $0.03$, we present some of the values of the PS in Table~\ref{table:4}). As stated earlier, when $\theta=\phi=51.827$ degrees, we get $q=q_{max}$ for NMES. So, to test the non-locality when $q=q_{max}$, we have to check whether $\hat{q}_{lb}>0$ as given in Equation~\eqref{hatqlb}. From Table~\ref{table:4}, we get $\bar{\Sigma}_{4}=0.0807$ and $S_{\Sigma_{4}}=0.0037$ which is value we get from the off-line phase as stated in Algorithm~\ref{alg1}. This value is constant for the pair of qubits $\left(Q_{3}, Q_{4}  \right) $ and will be different for other pairs. Now we will calculate the value of $\hat{q}_{lb}$.

From the experiment, we get $\bar{\epsilon}_{5}=0.1281$ with $S_{\epsilon_{5}}=0.0039$ and the values of CIs are given for $\left( 1-\alpha\right) \in \left\lbrace 0.99,0.95,0.90,0.80\right\rbrace $. To calculate  $\hat{q}_{lb}$ with different CIs, we will use Equation~(\ref{hatqlb}).  
From Table~\ref{table:6}, when  $\theta=\phi=51.827$ degrees, we get  $\hat{q}_{lb} > 0$ for different CIs. 
The error in estimating the value of $\hat{q}_{lb}$ is approximately $52\%$. If we increase the number of samples in the experiment, i.e., the value of $n$, then $\hat{q}_{lb}$ will also increase as seen in Figure~\ref{fig:7}. But despite the errors in the experiment, when $q=q_{max}$, we get the lower bound  on Hardy's probability greater than zero with $99\%$ confidence on the data. 

\subsubsection{Test of non-locality when $q < q_{max}$} \label{TestofNL_q<qmax}
From Table~\ref{table:6}, when $\theta=\phi=55$ degrees ($q=0.0886$), by a similar analysis, we get $\hat{q}_{lb}$ to be around $0.042$, for different CIs, i.e., $\hat{q}_{lb}>0$. The error in estimating the value of $\hat{q}_{lb}$, in this case, is around $52.5\%$.
Similarly, when $\theta=\phi=45$ ($q=0.0833$), the value of $\hat{q}_{lb}$ is around $0.02$. But the error in this case in estimating $\hat{q}_{lb}$ is around $76\%$. Clearly these results support a non-zero value of Hardy's probability.

But when $\theta=30$ and $\phi=60$ ($q=0.0433$), we estimate $\hat{q}_{lb}<0$. Similar result is obtained  when $\theta=60$ and $\phi=30$ ($q=0.0433$). So, form these results we cannot conclude that the state is really an NMES or not when $q=0.0433$.

When the value of $q$ is decreased further, when $q=0.00088$, the values of $\hat{q}_{lb}$ again become less than zero. So, form these results, we cannot conclude about the state as discussed above.
\subsubsection{Summary of the above two experiments} \label{summary_of_two_exp}
We know that in Hardy's test, we should get Hardy's probability $q > 0 $ for all NMES. But, from the above experimental data, we get for some NMES, the estimated lower bound on Hardy's probability $\hat{q}_{lb} \leq 0$. The mismatch between theoretical value and the estimated value from the experiments can be explained with respect to Equation~(\ref{q_bound}) as follows:
\begin{itemize}
\item for those NMES when $q$ is larger than $ \bar{\Sigma}_{4}$, the distinction between $\bar{\epsilon}_{5}$  and $\bar{\Sigma}_{4}$ is clear and $\hat{q}_{lb} > 0$.
\item But for those NMES when $q$ is less than $\bar{\Sigma}_{4}$, it is hard to distinguish between $\bar{\epsilon}_{5}$ and $ \bar{\Sigma}_{4}$, and we have $\hat{q}_{lb} \leq 0$.
\end{itemize}


\subsubsection{Consistency check} \label{ShiftQ3Q4}
\begin{figure}[h!]
\centering
\subfloat[$Q_{3}$ as control qubit and $Q_{4}$ as target qubit.]{\includegraphics[width = 0.47\linewidth]{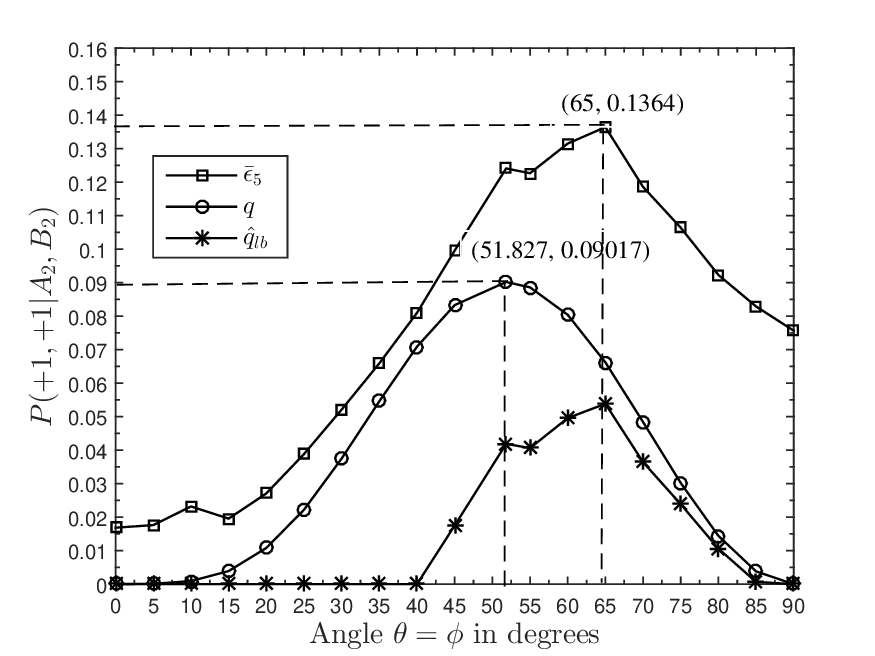}} \hspace{0.1cm}
\subfloat[$Q_{2}$ as control qubit and $Q_{1}$ as target qubit.]{\includegraphics[width = 0.47\linewidth]{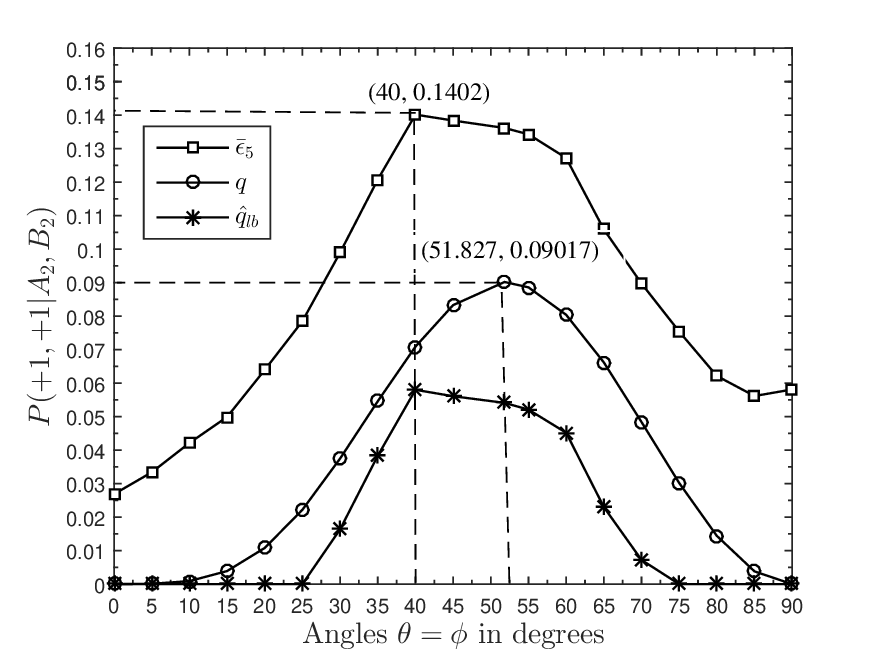}}
        \caption{The variation of $\bar{\epsilon}_{5}$, $q$, and $\hat{q}_{lb}$ against
$\theta=\phi$ for $(1-\alpha)=0.99$ and $n=10$.}\label{fig:8_9}
$\theta=\phi$]{The variation of $\bar{\epsilon}_{5}$, $q$, and $\hat{q}_{lb}$ against
$\theta=\phi$ for $(1-\alpha)=0.99$ and $n=10$ for (a) $Q_{3}$ as control qubit and $Q_{4}$ as target qubit, (b) $Q_{2}$ as control qubit and $Q_{1}$ as target qubit.}
\end{figure}

The above summary indicates that as $q$ decreases, so does $\bar{\epsilon}_{5}$. Thus it is a natural question to ask, whether at $q=q_{max}$, we get $\bar{\epsilon}_{5}=\bar{\epsilon}_{5}^{max}$ or not, where
\begin{equation}\label{Epsilon_5max}
\bar{\epsilon}^{max}_{5}=\max_{0 \leq \theta=\phi \leq 90} \lbrace \bar{\epsilon}_{5_{\theta=\phi}} \rbrace.
\end{equation} 
 We conduct another set of experiments to investigate this. If we take  $\theta=\phi$ and vary it from $0$ to $90$ degrees, a bell-shaped curve is obtained with a peak at $\theta=\phi=51.827$ degrees for $q$ as shown in Figure~\ref{fig:8_9} (a). For limited control of the IBM quantum computer, we plot $\bar{\epsilon}_{5} $,  $q$  and  $\hat{q}_{lb}$ with $\left( 1-\alpha\right)=0.99$ for  $\theta=\phi$ varying from $0$ to $90$ degrees with an increment of $5$ degrees, i.e.,  $\theta=\phi=5i$, where $i \in \lbrace 0,1,\ldots,18\rbrace$ (when $i=18$ we get $\theta=\phi=90$, then the value of $\chi$ is undefined as shown in the appendix. So we take $\theta=\phi=89.99$ for this case).

From Figure~\ref{fig:8_9} (a), we observe $ \bar{\epsilon}^{max}_{5} $, as defined in Equation~(\ref{Epsilon_5max}), is shifted right where $\theta=\phi=65$ degrees, instead of $\theta=\phi=51.827$ degrees. To get a more accurate result of $\bar{\epsilon}_{5}^{max}$, we repeat this experiment for $75 \geq \theta=\phi \geq 55$ degrees with an increment of one degree. The result shows that $\bar{\epsilon}_{5}^{max}$ occurs where $\theta=\phi=62$ degrees (not shown in Figure~\ref{fig:8_9} (a)). This procedure can be repeated again  if the more precise occurrence of $\bar{\epsilon}_{5}^{max}$ is needed. We also observe that $\hat{q}_{lb}>0$ when $85 \geq \theta=\phi \geq 40$ degrees and zero elsewhere, whereas $q > 0$ for $90 > \theta=\phi > 0$ degrees. This mismatch happens because of the reasons stated in Section~\ref{summary_of_two_exp}. From these results it can be concluded that although in the IBM quantum computer, we get a non-zero value of the lower bound on Hardy's probability for NMES, but the errors occurred in the computer need to be more stable.


\subsection{Other possible combinations of multi-qubit gate} \label{Other_CNOT_Gate}
In this section, we check Hardy's non-locality for the rest of the available combinations of qubits in the IBM quantum computer.

 \begin{table} [h!]
\centering
 \caption{The values of $\bar{\epsilon}_{5}$ for $n=10$ for all possible pairs of qubits other than $\left( Q_{3},Q_{4}\right)$ when $\theta=\phi \in \left\lbrace 45,51.827,55\right\rbrace $ in degrees.}  \label{table:7}
\begin{tabular}{ p{1.2cm}   p{1cm}  p{1.5cm}  p{1.5cm}   p{1.5cm}}
 \hline   Control  & Target & \multicolumn{3}{c}{values of $\bar{\epsilon}_{5}$ for $n=10$ for $\theta=\phi$} \\
\cline{3-5} qubit & qubit & \hspace{4mm} 45 &  \hspace{1mm} 51.827 &  \hspace{4mm} 55\\
 \hline 
  $Q_{2}$ & $Q_{0}$   &0.101022 &  0.111367 & 0.116217 \\

  $Q_{3}$ & $Q_{2}$   & 0.094067  & 0.112275 &   0.114400 \\
 $Q_{1}$ & $Q_{0}$   & 0.099931  & 0.119311 & 0.119497 \\

 $Q_{2}$ & $Q_{1}$   & 0.138400 & 0.136175 & 0.134289
 \\

 $Q_{2}$ & $Q_{4}$   & 0.138505 & 0.150186 & 0.157253\\
\hline
\end{tabular}
\end{table}
\subsubsection{Check for non-locality}
There are currently six combinations of multi-qubit gate implementation available  in \textit{ibmqx4}~\cite{IBM}. For the multi-qubit $CNOT$ gate that we use, the possible control qubit and target qubit pairs other than $\left( Q_{3},Q_{4}\right)$, are summarized in Table~\ref{table:7}. We measure $\bar{\epsilon}_{5}$ for $\theta=\phi \in \left\lbrace 45,51.827,55\right\rbrace $ degrees with all combinations of pairs of qubits. It can be seen, when $\theta=\phi=51.827$ degrees, the value of $\bar{\epsilon}_{5}$ is minimum for the pair $\left( Q_{2},Q_{0}\right)$ and maximum for the pair $\left( Q_{2},Q_{4}\right)$. All the experiments described earlier have been performed using all combinations of these pair of qubits as described in Table~\ref{table:7} and similar conclusions of the non-locality are obtained as we get for the pair $\left( Q_{3},Q_{4}\right)$.

\subsubsection{Consistency check and shift of the peaks} \label{Shiftother}
We want to see, for other possible two-qubit pairs, whether $ \bar{\epsilon}_{5}^{max} $ (as defined in Equation~(\ref{Epsilon_5max})) occurs when $q=q_{max}$  or not. We also want to see, in case there is any shift, whether it is in the same direction as shown in Figure~\ref{fig:8_9} (a) for the pair $\left( Q_{3},Q_{4}\right)$ or not.

From Table~\ref{table:7}, it can be seen for the pair $\left( Q_{2},Q_{1}\right)$, when $\theta=\phi=51.827$ degrees, the value of $\bar{\epsilon}_{5}$  is less than the  value when $\theta=\phi=45$ degrees and greater than the value when $\theta=\phi=55$ degrees, i.e., $ \bar{\epsilon}_{5_{\theta=\phi=45}}>\bar{\epsilon}_{5_{\theta=\phi=51.827}} > \bar{\epsilon}_{5_{\theta=\phi=55}} $. To verify this result, we perform a similar experiment for the pair $\left( Q_{2},Q_{1}\right)$ as we did  for the pair $\left( Q_{3},Q_{4}\right)$ (as shown in Figure~\ref{fig:8_9} (a)). Results are shown in Figure~\ref{fig:8_9} (b) which indicates that there is a shift of the value of $\bar{\epsilon}_{5}^{max}$ to the left for the pair $\left( Q_{2},Q_{1}\right)$ and it occurs when $\theta=\phi=40$ degrees. 
For the rest of the pairs, we find that  $\bar{\epsilon}_{5}^{max}$ is shifted to the right, i.e., $ \bar{\epsilon}_{5_{\theta=\phi=55}}>\bar{\epsilon}_{5_{\theta=\phi=51.827}} > \bar{\epsilon}_{5_{\theta=\phi=45}} $ as shown in Table~\ref{table:7}. So, we can conclude that $\bar{\epsilon}^{max}_{5}$ did not occur at $q=q_{max}$, rather it is shifted to the right or to the left due to the errors.

\subsection{Application of the shift of the peaks in quantum  protocols using Hardy's test}
For some of the protocols like quantum  Byzantine agreement~\cite{QBA}, it is necessary to check whether Hardy's state (the state for which Equations~\eqref{eq:1}-\eqref{eq:4} are satisfied) is actually prepared or not. Suppose, theoretically  $q=q_{max}$ is achieved for a specific value $\rho$ of the relevant parameter (of which $q$ is a function, like $\theta$ and $\phi$ in our experiment). Because of the peak shifts as described in Section~\ref{ShiftQ3Q4} and~\ref{Shiftother}, in practical experiments, $q_{max}$ should not be estimated corresponding to the exact value of $\rho$, but in an interval around $\rho$. 

During the experiment, let the value of $q_{max}$ with the addition of errors be $Q_{max}$ (like $\bar{\epsilon}^{max}_{5}$ in our experiment). Now if the errors in the experiment are not stable, it is expected that $Q_{max}$ will lie in an interval around $\rho$, i.e., $\left\lbrace \rho-\delta,\rho+\delta \right\rbrace$ for some $\delta > 0$. For example, in our experimental set-up, $\delta$ is found to be $12$ degrees when $\rho=\left( 51.827, 51.827 \right) $ in degrees of the parameter  $ \left( \theta , \phi \right) $.  

\subsection{Verifying whether reducing the number of gates reduces the error in the circuit}
\begin{figure}[h!]
\centering
\subfloat[Quantum circuit for the PS $\ket{\psi} = \frac{1}{\sqrt{2}}(\ket{0} + \ket{1}) \otimes \ket{0}$ when  $\theta=\phi=0$ degrees.]{\includegraphics[width = 0.35\linewidth]{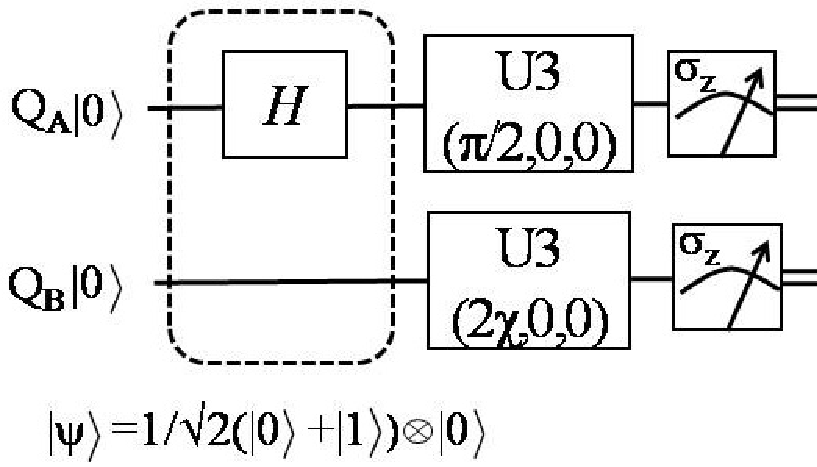}} \hspace{1cm}
\subfloat[Quantum circuit for the PS $|\psi\rangle = \frac{1}{\sqrt{2}}(|0\rangle + |1\rangle) \otimes |1\rangle $ when  $\theta=90$ and $\phi=0$ degrees.]{\includegraphics[width = 0.35\linewidth]{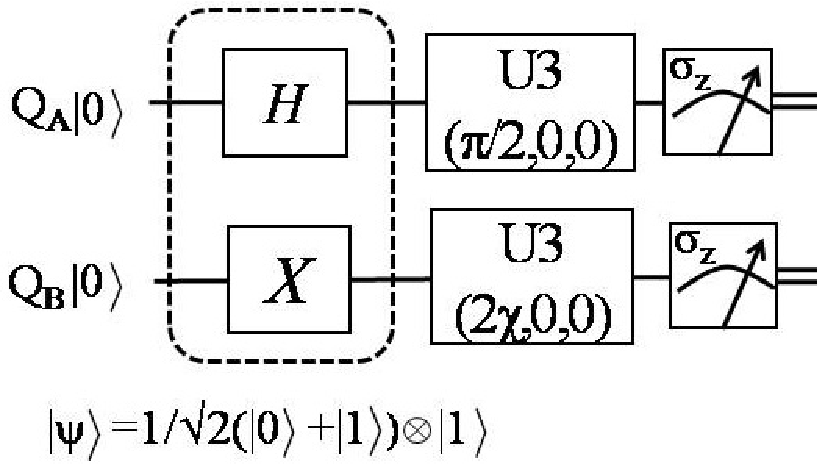}}
        \caption{Quantum circuit and measurement for $P(+1,+1|A_{1},B_{2})$ when number of gates are reduced significantly. }\label{fig:10}
\end{figure}

To verify whether reducing the number of gates in the circuit reduces the error or not, we perform another series of experiments for $\left( Q_{3},Q_{4}\right)$ pair of qubits. For  $\theta=\phi=0$, we get a PS, i.e., $|\psi\rangle = \frac{1}{\sqrt{2}}(|0\rangle + |1\rangle) \otimes |0\rangle $. This state can be created easily by using a Hadamard gate $H$ in Alice's qubit and $U_{1}$ and $U_{3}$ used  in Figure~\ref{fig:5}, becomes the identity ($Id$) gate. So, the number of gates is reduced significantly. For the modified circuit as shown in Figure~\ref{fig:10} (a), the value of $\bar{\epsilon}_{4}=0.0084$ for $n=10$ is less than the previous value (as shown in Table~\ref{table:4}), i.e., $ 0.0193 $ 

Also, for $\theta=90,\phi=0$,  we get a PS $|\psi\rangle = \frac{1}{\sqrt{2}}(|0\rangle + |1\rangle) \otimes |1\rangle $. For this state, a Hadamard gate on Alice's qubit and a bit flip gate $X$ on Bob's qubit is required as shown in Figure~\ref{fig:10} (b). Experimental results show that $\bar{\epsilon}_{4}=0.0079$ for $n=10$ is again less than the previous value (as shown in Table~\ref{table:4}), i.e., $0.0209$.

These experiments are repeated for the rest of the pairs of qubits and similar results are obtained. We can conclude that reducing the number of gates reduces the error in the circuit of the IBM quantum computer. 
 
\subsection{ Study of change of errors with respect to time in superconducting qubits}
 \begin{table} [h!]
\centering
\caption{The values of $\bar{\epsilon}_{5}$ for $n=10$ for  the pair $\left( Q_{3},Q_{4}\right)$ before and after one month when $\theta=\phi=51.827$ in degrees, SD=standard deviation, and CI=confidence interval.} \label{table:8}
\begin{tabular}{p{1.25cm} p{1.25cm}   p{1.25cm}  p{1.25cm}  p{1.25cm}   p{1.25cm}  p{1.25cm}}
 \hline  Time  & $\bar{\epsilon}_{5}$ & SD & \multicolumn{4}{c}{ $t_{\frac{\alpha}{2}}\frac{S_{\epsilon_{5}}}{\sqrt{n}}$ for different CIs} \\
 \cline{4-7}line & &($ S_{\epsilon_{5}}$) & \hspace{3mm} $99\%$ & \hspace{3mm} $95\%$  &  \hspace{3mm} $90\%$  &  \hspace{3mm} $80\%$    \\
 \hline 
 Before & 0.16254 & 0.0078 & 0.154524 &	0.15696 &	0.158018 &	0.159129 \\
 After & 0.1281 &	0.0039 & 0.124092 & 0.12531	& 0.125839	& 0.126394 \\
\hline
\end{tabular}
\end{table} 
During the experiments, the IBM quantum computer has undergone maintenance for nearly one month when some of the experiments were done. To present all the results in the same time-line, we  repeat all the previous experiments. When we compare the data of one month earlier experiments, we find a significant change of values, similar to what was reported in~\cite{mermin2016}. For  the pair $\left( Q_{3},Q_{4}\right)$ when $\theta=\phi=51.827$ degrees, one month earlier, we got the value of $\bar{\epsilon}_{5}=0.16254$, $S_{\epsilon_{5}}= 0.0078$ and the values of $t_{\frac{\alpha}{2}}\frac{S_{\epsilon_{5}}}{\sqrt{n}}$ for different values of CIs are given in Table~\ref{table:8}. From this, we can see that the value of $\bar{\epsilon}_{5}$ one month earlier is greater than the values of what we get one month later. A similar trend is noticed for the rest of the data.

\subsection{Benchmarking of superconducting quantum devices using Hardy's paradox}
Benchmarking is the process by which the performance of any computing device is evaluated. In the current era of Noisy Intermediate-Scale Quantum (NISQ)~\cite{Preskill2018}, choosing a universal metric for benchmarking is extremely difficult, because computing can be performed using various quantum technologies such as 
 superconducting qubits~\cite{Gambetta2017,Krantz2019}, ion trap~\cite{HAFFNER2008}, optical lattices~\cite{Treutlein2006}, quantum dots~\cite{Loss1998}, nuclear magnetic resonance (NMR)~\cite{lu2016nmr}, etc. Also, how the noise affects the devices and their characterization are not well-understood.

Researchers have proposed different metrics for benchmarking, for instance, fidelity~\cite{Beale2018}, unitarity~\cite{Wallman2015}, quantum volume~\cite{Cross2019}, quantum chemistry~\cite{McCaskey2019}, etc., each having their own drawbacks~\cite{1912.00546}. In~\cite{mermin2016}, the authors have also demonstrated non-locality in the case of Mermin polynomials for three, four and five qubits. They have concluded that the fidelity of the quantum computer decreases when the number of qubits is increased from three. However, they do not mention anything for two qubits. From the experimental test of Hardy's paradox, we propose two metrics for benchmarking of two qubits of any superconducting quantum device as discussed below.

First, from Section~\ref{TestofNL_q=qmax} and~\ref{TestofNL_q<qmax}, it can be seen that, for those states where $q > \bar{\Sigma}_{4}$,  we get $\hat{q}_{lb} > 0$ which supports a non-zero value of Hardy's probability. When $\bar{\epsilon}_{4}^{max}< q <q_{max}$, the previous conclusion is still valid.  But for those states where  $q < \bar{\Sigma}_{4}$, no conclusion can be drawn about the value of  Hardy's probability as $\hat{q}_{lb} \leq 0$. So, the minimum value of $q$ from which we get $\hat{q}_{lb} > 0$ (as illustrated in Table~\ref{table:6}), can be the performance measure of a quantum computer. Lesser the value of $q$ which supports $\hat{q}_{lb} > 0$, the better the performance of the quantum computer. This parameter measures how well the device realizes NMES states.

Second, although we get $q_{max}$ at $\theta=\phi=51.827$ degrees, during the experiment, due to unstable errors, the value of $\bar{\epsilon}_{5}^{max}$ may be shifted to any nearby value, as shown in Section~\ref{ShiftQ3Q4} and~\ref{Shiftother}. In our case, it shifted to the left for the pair $\left( Q_{2},Q_{1}\right)$ and to the right for the rest of the pairs of qubits. For our experiment when $\theta=\phi$, the amount of shift is $12$ degrees. So, the amount of shift can be considered as a performance measure of a quantum computer. The smaller the shift, the better is the performance. This parameter measures the precision of the quantum device.

\section{Conclusion and Future Work} 
We have performed an experimental verification of Hardy's paradox of non-locality for the first time in superconducting circuits. 
Our initial motivation was to check it for two qubits in the \textit{ibmqx4} five-qubit chip, by choosing any two from the five qubits. 
When $\Sigma_{4} < q \leq q_{max}$, the estimated lower bound $\hat{q}_{lb}$ on Hardy's probability is found to be greater than zero, supporting non-locality. But when $ q \leq \Sigma_{4}$, we get  $\hat{q}_{lb} \leq 0$, because then the errors become of the same order as $q$. 

Interestingly, though $\bar{\epsilon}_{5}^{max}$ decreases with $q$, experimental results show that
$\bar{\epsilon}_{5}^{max}$ does not occur at $q=q_{max}$, rather we get a shift of $\bar{\epsilon}_{5}^{max}$ to the right for the $\left( Q_{3},Q_{4}\right)$ pair of qubits. We also show that the shift direction is not constant, whether it is to the right or left depends on the pair of qubits.

Moreover, we have shown that the above type of shift can occur during the practical implementation of any Hardy's paradox based quantum protocols like quantum Byzantine agreement and we have also discussed possible remedies.

Based on the results of our experiments, we have proposed two performance measures of any quantum computer for two qubits. First, the minimum value of $q$ above which non-locality is established. Second, the amount of shift needed to get the experimental maximum value of $\bar{\epsilon}_{5}^{max}$ of Hardy's probability.

Further, we have performed experiments to show how decreasing the number of gates in the circuits decreases the errors in the circuit for all possible pairs of qubits.

We have also studied the change of errors in IBM quantum computer over time and concluded that errors are decreasing over time.

From the theoretical analysis of the Hardy's experimental set-up, we have found that this test fails for all NMES, where the value of $\phi=90$ and $\theta \neq \left\lbrace 0,45,90\right\rbrace $ degrees. Future work may consider the possibility of a new test for Hardy's paradox for two qubits, so that it does not fail for any NMES.

\appendix
\section*{Appendix}
\subsection*{Boundary Values of $\chi$}
\label{app1}
In Section \ref{Circuits for Hardy's equations}, $\chi$  is defined as $\text{cot} \chi = \text{tan} \theta \text{cos} \phi$. When $\theta = 90$ degree and $\phi= 90$  degree, we get $\text{tan} \theta = \infty $ and $\text{cos} \phi = 0$, which leads to  $\text{cot} \chi = \infty \cdot 0$. This is an indeterminate form. 

When $\theta \to 90+$ and $\phi \to 90+$, the value of $\chi$ is positive.
When $\theta \to 90+$ and $\phi \to 90-$, the value of $\chi$ is negative.
When $\theta \to 90-$ and $\phi \to 90+$, the value of $\chi$ is negative.
When $\theta \to 90-$ and $\phi \to 90-$, the value of $\chi$ is positive.

So, the limit does not exist. More formally, 
$$\lim_{\left(\theta, \phi \right) \to \left( \frac{\pi}{2}, \frac{\pi}{2} \right)} \text{tan} \theta \text{cos} \phi =\lim_{\left( x, y \right) \to \left( 0, 0 \right)} \text{tan} \left( x +\frac{\pi}{2}\right)  \text{cos} \left( y + \frac{\pi}{2}\right)
=\lim_{\left( x, y \right) \to \left( 0, 0 \right)} \text{cot} x \text{sin} y,$$
where $x=\left(\theta-\frac{\pi}{2} \right) $ and  $y=\left(\phi-\frac{\pi}{2} \right) $.
Now changing the co-ordinate system from rectangular to polar co-ordinate system and substituting $x=r$ cos$\varphi$ and $y=r$ sin$\varphi$, we can write the above expression as 

\begin{eqnarray*}
& & \lim_{ r  \to 0 }  \text{cot}\left(  r \text{cos} \varphi \right) \text{sin} \left( r \text{sin} \varphi\right)\\
&=&\lim_{r  \to 0 } \frac{\text{cos} \left(  r \text{cos} \varphi \right)}{\text{sin} \left(  r \text{cos} \varphi \right)} \text{sin} \left( r \text{sin} \varphi\right) = \lim_{r  \to 0 }  \frac{\left( 1 - \left( r \text{cos}  \varphi\right)^{2} + \cdots \right) }{\left( r \text{cos} \varphi - \frac{\left(r \text{cos} \varphi ^{3} \right)}{3!} + \cdots  \right)}{\left( r \text{sin} \varphi - \frac{\left(r \text{sin} \varphi ^{3} \right)}{3!} + \cdots  \right)}\\
&=&\lim_{r  \to 0 }  \frac{\left( 1 - \left( r \text{cos}  \varphi\right)^{2} + \cdots \right) }{r \left(  \text{cos} \varphi - \frac{r ^{2}\left( \text{cos} \varphi ^{3} \right)}{3!} + \cdots  \right)}{r \left(  \text{sin} \varphi - \frac{r^{2}\left( \text{sin} \varphi ^{3} \right)}{3!} + \cdots  \right)} = \frac{\text{sin} \varphi}{\text{cos} \varphi}=\text{tan} \varphi.
\end{eqnarray*}
 So, $\chi$ doesn't have any definite value, but it depends on $\varphi$.


\begin{thebibliography}{0}
\bibitem{ibmqx4} IBM has several chips and a subset of those becomes available for experiments from time to time. Currently, the ibmqx4
chip is under maintenance, while the available chips are ibmqx2, ibmq vigo, ibmq ourense and ibm 16 melbourne.
\bibitem{Matlab} We use MATLAB\textsuperscript{\textregistered} to get these values.
\bibitem{onesided} Two-sided test typically applies to $\mathcal{H}_{0}: X =p$ vs $\mathcal{H}_{1}: X \neq p$ and one-sided test $\mathcal{H}_{0}: X = p$ vs $\mathcal{H}_{1}: X > p$ or $\mathcal{H}_{1}: X < p$ where $X$ is the chosen parameter under test and $p$ is a specific value.
\end{thebibliography}

\end{document}